\documentclass[twocolumn]{aastex62}
\pdfoutput=1 
\usepackage{amsmath,amstext}
\usepackage[T1]{fontenc}
\usepackage{apjfonts} 
\usepackage[figure,figure*]{hypcap}
\usepackage{natbib}
\usepackage[disable,colorinlistoftodos]{todonotes}
\usepackage{xcolor}
\usepackage{soul, CJK}

\let\Oldtodo\todo
\renewcommand{\todo}[1]{\Oldtodo[inline]{#1}}

\shorttitle{ZTF Science Objectives}
\shortauthors{Graham et al.}

\begin{document}
\begin{CJK*}{UTF8}{bkai}

\title{The Zwicky Transient Facility: Science Objectives}

\author{Matthew J.\ Graham}
\affiliation{Division of Physics, Mathematics and Astronomy, California Institute of Technology, Pasadena, CA 91125, USA}
\email{Corresponding author: mjg@caltech.edu}

\author{S.\ R.\ Kulkarni}
\affiliation{Division of Physics, Mathematics and Astronomy, California Institute of Technology, Pasadena, CA 91125, USA}

\author[0000-0001-8018-5348]{Eric C.\ Bellm}
\affiliation{DIRAC Institute, Department of Astronomy, University of Washington, 3910 15th Avenue NE, Seattle, WA 98195, USA}

\author[0000-0001-5855-5939]{Scott M.\ Adams}
\affiliation{Division of Physics, Mathematics and Astronomy, California Institute of Technology, Pasadena, CA 91125, USA}

\author{Cristina Barbarino}
\affiliation{The Oskar Klein Centre \& Department of Astronomy, Stockholm University, AlbaNova, SE-106 91 Stockholm, Sweden}

\author[0000-0003-0901-1606]{Nadejda Blagorodnova} 
\affiliation{Division of Physics, Mathematics and Astronomy, California Institute of Technology, Pasadena, CA 91125, USA}

\author[0000-0002-2668-7248]{Dennis Bodewits} 
\affiliation{Department of Astronomy, University of Maryland, College Park, MD 20742, USA}
\affiliation{Department of Physics, Auburn University, Auburn, AL 36849, USA}

\author[0000-0002-4950-6323]{Bryce Bolin}
\altaffiliation{B612 Asteroid Institute and DIRAC Institute Postdoctoral Fellow}
\affiliation{Department of Astronomy, University of Washington, Seattle, WA 98195, USA}
\affiliation{B612 Asteroid Institute, 20 Sunnyside Ave, Suite 427, Mill Valley, CA 94941}

\author[0000-0002-4611-9387]{Patrick R.\ Brady}
\affiliation{Center for Gravitation, Cosmology and Astrophysics, Department of Physics, University of Wisconsin--Milwaukee, P.O. Box 413, Milwaukee, WI 53201, USA}

\author[0000-0003-1673-970X]{S.\ Bradley Cenko}
\affiliation{Astrophysics Science Division, NASA Goddard Space Flight Center, MC 661, Greenbelt, MD 20771, USA}
\affiliation{Joint Space-Science Institute, University of Maryland, College Park, MD 20742, USA}

% ...


\begin{abstract}
The Zwicky Transient Facility (ZTF), a public-private enterprise, is a new time domain survey employing a dedicated camera on the Palomar 48-inch Schmidt telescope with a 47\,deg$^2$ field of view and 8 second readout time. It is well positioned in the development of time domain astronomy, offering operations at 10\% of the scale and style of the Large Synoptic Survey Telescope (LSST) with a single 1-m class survey telescope. 
The public surveys will cover the observable northern sky every three nights in $g$ and $r$ filters and the visible Galactic plane every night in $g$ and $r$. Alerts generated by these surveys are sent in real time to brokers. A consortium of universities which provided funding (``partnership") are undertaking several 
boutique surveys. The combination of these surveys producing one million alerts per night allows for exploration of transient and variable astrophysical phenomena brighter than $r \sim 20.5$ on timescales of minutes to years.  We describe the primary science objectives driving ZTF including the physics of supernovae and relativistic explosions, multi-messenger astrophysics, supernova cosmology, active galactic nuclei and tidal disruption events, stellar variability, and Solar System objects.
\end{abstract}



\section{Introduction}
The past decade has seen an explosion in time domain astronomy driven by the availability of new instruments and facilities dedicated to repeated observations of large areas of sky. A number of surveys, e.g., Catalina Real-Time Survey
(CRTS; \citealt{Drake2009}), 
Palomar Transient Factory (PTF/iPTF; \citealt{Law:09:PTFOverview}),
Panoramic Survey Telescope and Rapid Response System (Pan-STARRS or PS; \citealt{Kaiser2004}), 
All Sky Automated Survey for SuperNovae (ASAS-SN; \citealt{Shappee2014}), 
The Asteroid Terrestrial-impact Last Alert System (ATLAS; \citealt{Tonry2018}), have opened up the exploration of temporal behavior from Solar System objects to variable stars in the Galaxy to relativistic explosions across the universe. They have each employed differing modes of operation, e.g., the number of repeat visits to the same region of sky per night, inter-nightly cadence, choice of filters, etc., in addition to the varying capabilities of camera, telescope, and site to probe the potential discovery space. 

The Zwicky Transient Facility (ZTF) consists of a wide-field imager on the
Palomar 48-inch Oschin (Schmidt) telescope and an integral field unit spectrograph (IFUS)
on the Palomar 60-inch telescope 
optimized for spectral classification of relatively bright ($<19\,$mag) explosive transients. The former likely represents the height of what can be achieved with a single 1-meter class survey telescope. The resulting transients and variables will be bright enough that followup can be undertaken for well-defined samples by the existing suite of larger telescopes. 

ZTF operates in a rich landscape of optical time domain surveys. ASAS-SN is a long term project dedicated to surveying the entire night sky. It consists of telescopes in both hemispheres and at a number of longitudes and is well positioned to survey the entire night sky to about 18.5\,mag each night. PS-1 has
already delivered the definitive photometric catalog and deep reference imaging for the northern sky. PS-1 is now largely dedicated to the study of Near-Earth Asteroids (NEOs).  ATLAS visits a significant portion of the night sky every two nights, to about 19.5\,mag. Although ATLAS is also tuned for investigating near-earth asteroids, it is reporting transients and publishing light curves for variable stars.  

ZTF is undertaking a number of of different surveys (with cadence ranging from minutes to days).
Its primary strength is its combination of depth and areal survey speed \citep{2016PASP..128h4501B}, which enable it to identify transients at earlier times relative to ASAS-SN and ATLAS while covering a large fraction of the accessible night sky. Its productivity in the discovery of transients is enhanced by the availability of dedicated spectroscopic followup to provide routine classification.
The IFUS on the robotic 60-inch telescope can undertake, without human
intervention, nearly a dozen spectral
classification observations per night.  Finally,
ZTF has been designed to be a stepping stone for LSST
for transient (and variable) object astronomy. In particular, real time
alerts\footnote{We use the LSST definition of an alert: a 5$\sigma$ change
in RA, Declination or flux, with respect to the reference sky. The resulting alert stream provides a comprehensive, real-time inventory of everything ZTF knows about the changing night sky, including not just transients but also variable stars and solar system objects.} are being
issued to brokers to enable community use of the data and to develop infrastructure for the LSST era. Although the main focus of ZTF (given its name) may initially appear to be transient science, it will also contribute to our knowledge and understanding of optical phenomena in both real time and archival behaviors.


In this paper, we describe the science objectives that motivated the ZTF project. These are defined around particular areas of interest and we present here the expected outcomes and science deliverables of each: the physics of supernovae (SNe) and relativistic explosions (sec. 3), multi-messenger astrophysics (sec. 4), cosmological distances from Type Ia SNe (sec. 5), cosmology with gravitationally lensed SNe (sec. 6), active galactic nuclei (AGN) and tidal disruption events (TDEs) (sec. 7), Galactic science (sec. 8), small Solar System bodies (sec. 9), and astroinformatics and astrostatistics (sec. 10). 

The following papers discuss performance and sub-systems in detail:
\citet{tmp_Bellm:18:Overview} give a general overview of the ZTF system.
\citet{tmp_Dekany:18:ZTFObservingSystem} provide an in-depth description of the design of the observing system.
\citet{tmp_Bellm:18:ZTFScheduler} discuss the ZTF surveys and scheduler.
\citet{tmp_Masci:18:ZTFDataSystem} detail the ZTF data system.
\citet{tmp_Patterson:18:ZTFAlertDistribution} present the alert distribution system employed by ZTF.
\citet{tmp_Mahabal:18:ZTFMachineLearning} discuss several applications of machine learning used by ZTF.
\citet{tmp_Tachibana:18:PS1StarGalaxy} present a new star/galaxy classifier developed for ZTF from the Pan-STARRS DR1 catalog \citep{Chambers:16:PS1,Flewelling:16:PS1db}.
\citet{tmp_Kasliwal:18:GROWTHMarshal} describe a web-based interface used by the ZTF collaboration to identify, track, and follow up transients of interest.

\section{The Zwicky Transient Facility}

ZTF employs a new 576 megapixel camera \citep{Dekany2016} with a 47\,deg$^2$ field of view on the Samuel Oschin 48" Schmidt telescope (P48) at Palomar Observatory. It can observe 3760 deg$^2$ per hour to a $5 \sigma$ detection limit of 20.5\,mag in {\it $r$} (with a 30\,s exposure). 
The data processing pipelines are managed by IPAC \citep{tmp_Masci:18:ZTFDataSystem} with different 
branches for single epoch images and catalogs, image subtraction, and moving objects. Alerts from difference images generated using the ZOGY algorithm \citep{Zackay:16:ZOGY} are produced within 20 minutes of the raw image being taken with the ZTF instrument and distributed using the Kafka system \citep{tmp_Patterson:18:ZTFAlertDistribution} operated by the University of Washington. These data packets contain thumbnails of the discovery, reference, and difference images as well as a 30-day light curve history for an alert.

Due to its funding profile, ZTF operates a unique observing strategy: 40\% of the time is for public surveys, 40\% for partnership observations, and the remaining 20\% for Caltech programs. For the public surveys \citep{tmp_Bellm:18:ZTFScheduler}, the entire visible sky from Palomar is observed in $g$ and $r$ every three nights and the visible Galactic Plane ($|b| < 7^\circ$) covered in $g$ and $r$ every night. Alerts from the public survey data are issued in real time. However, images and catalogs remain embargoed to all parties until public data releases (the first is scheduled for Spring 2019 and thereafter every semester).

ZTF is issuing of order one million alerts per night. 
Note that other next generation surveys can also produce alerts at this scale but do not currently make them public. 
The 3-year duration of the project also means that about 1 billion sources will be observed about one thousand times. The final table of all individual source detections will thus be a trillion row catalog. ZTF can be regarded to be a precursor to LSST, operating at roughly the 10\% scale (see Table~\ref{table1} for further comparisons between ZTF, LSST, and other surveys). 

\begin{table*}
\centering
\caption{A comparison between ZTF, LSST, and other next generation surveys in terms of scale.}
\label{table1}
\begin{tabular}{lccccc}
\hline
Category & ATLAS & ASAS-SN & Pan-STARRS & ZTF & LSST \\
\hline
Number of total sources & - & $1 \times 10^8$ & $1 \times 10^{10}$ & $1 \times 10^9$ & $37 \times 10^9$ \\
Number of total detections & $1 \times 10^{12}$ & $1 \times 10^{11}$ & $1 \times 10^{11}$ & $1 \times 10^{12}$ & $37 \times 10^{12}$ \\
Annual visits per source & 1000$^c$ & 180$^d$ & 60$^e$ & 300$^a$ & 100$^b$ \\
Number of pixels & $1 \times 10^8$ & $4 \times 10^6$ (x 4) & $1 \times 10^9$ & $6 \times 10^8$ & $3.2 \times 10^9$ \\
CCD surface area (cm$^2$) & 90 & 9 & 1415 & 1320  & 3200  \\
Field of view (deg$^2$) & 30 & 4.5 & 7 & 47 & 9 \\
Hourly survey rate (deg$^2$) & 3000 & 960 & - & 3760 & 1000 \\
$5 \sigma$ detection limit in $r$ & 19.3 & 17.3 & 21.5 & 20.5 & 24.7 \\
Nightly alert rate & - & - & - & $1 \times 10^6$ & $1 \times 10^7$ \\
Nightly data rate (TB) & 0.15 & - & - & 1.4 & 15 \\
Telescope (m) & 0.5 & $4 \times 0.14$ & 1.8 & 1.2 & 6.5 \\
No. of telescopes & 2 (6) & 5 & 2 & 1 & 1 \\
\hline
\end{tabular}

$^a$ - in 3 filters; $^b$ - in 6 filters; $^c$ - in 2 filters; $^d$ - in 2 filters; $^e$ - in 5 filters
\end{table*}

A major motivation of ZTF was the detection and study of infant explosions.To this end, ZTF  also has a dedicated follow-up capability in the form of the  Spectral Energy Distribution Machine (SEDM) on the Palomar 60'' telescope \citep{Blagorodnova18}. This combines a low resolution (R $\sim$ 100) integral field unit (IFU) spectrograph with a multi-band $(ugri)$ photometer and is optimized for classification and high observing efficiency. Sources detected by ZTF can be (automatically) submitted to the SEDM observing queue for swift observation. Below we discuss each of the main science areas that ZTF is expected to explore. As with other surveys in the past, we also anticipate that the wealth of new data provided by ZTF will enable serendipitous discoveries of new classes of rare events (e.g.,\ AT2018cow/ATLAS18qqn).

\section{Physics of supernovae and relativistic explosions}

Supernovae will be the major class of non-moving transients detected by ZTF and the expected rates across the range of different types of supernova support a number of systematic studies into these phenomena. 

\subsection{The quest and study of infant supernova explosions}
\label{sec:InfantSN}

One of the boutique surveys carried out by ZTF is high cadence (6 times a night) observations of 
two thousand square degrees of the sky. This survey with good depth (20.5\,mag) and good cadence was designed explicitly to find young SNe and undertake rapid follow up studies.  For massive star explosions, very early photometry from the ground and from space, available only for a handful of serendipitously observed events so far (e.g., \citealt{Campana06,AMS08,Gezari08,Garnavich16,Bersten18}) probes the early physics of explosion shock breakout and cooling (see \citealt{Waxman17} for a recent review). For massive star explosions, early photometry (especially if it includes space UV data, e.g., \citealt{Ofek10,Gezari08,Yaron17,Ganot16,Rubin17}) provides powerful constraints on the nature of the progenitor and the parameters of explosion (e.g., its energy per unit mass, \citealt{Rubin16}). 

For Type Ia SNe, early photometry, especially in the UV, is a powerful probe for the existence of a possible mass-donor companion to the exploding white dwarf \citep{Kasen10}. Initial reports about a handful of events (e.g., \citealt{Cao15,GH17}) motivate further exploration of this approach. Together, early photometric studies of both core-collapse and Type Ia SNe motivate a strong synergistic program combining ZTF data with rapid response {\it Swift} UV observations. 

\begin{figure}
\vspace{-3cm}
\hspace*{-1cm}\includegraphics[width=10.5cm]{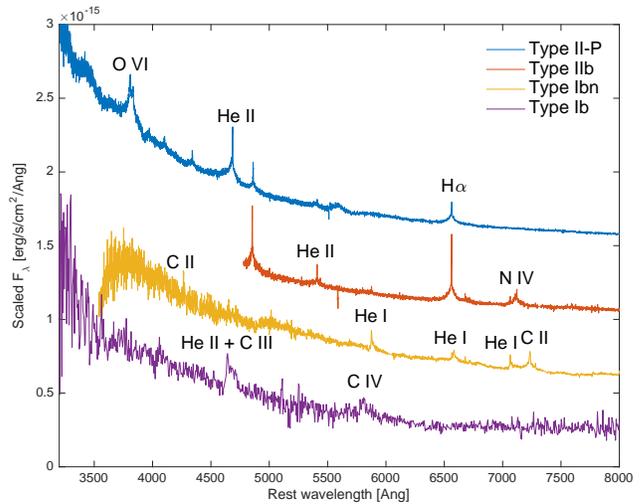}
\vspace{-3.5cm}
\caption{A collection of flash spectra from iPTF, showing the differing composition of the CSM around core-collapse SNe.}
\label{Figflash}
\end{figure}

Discovery of SNe within 24 hours of explosion, enabled by the ZTF discovery, coupled with the ability to rapidly
trigger the SEDM and other follow-up resources, would allow Target of Opportunity (ToO) spectroscopy
of young SNe within hours of explosion. As shown by initial results using this ``flash spectroscopy'' technique on iPTF triggers (e.g., \citealt{AGY+14,Khazov16,Yaron17}), analysis of such early spectra of massive star explosions allows us to extract unique information about the distribution of circumstellar material (CSM) around exploding stars. The composition of such material, measured from emission line intensities, provides a direct measurement of the surface composition of the supernova progenitor as it was prior to explosion, while the spatial distribution of the CSM, revealed by the transient nature of the emission lines, provides a record of the stellar mass loss just prior to explosion, with potentially critical clues about the SN explosion mechanism \citep{AGY+14,Yaron17}. Temporal evolution of the emission lines within a night \citep{Yaron17} provides a measurement of the temporal evolution of the temperature in the emitting material, a valuable constraint on shock and interaction physics. Mapping the properties of flash-spectroscopy-revealed CSM across the range of SN types (Fig.~\ref{Figflash}) is a key goal of ZTF. 

\subsection{New insights into interacting supernovae}
\label{sec:InteratingSNe}

Type IIn supernovae are SNe whose spectra show bright and narrow ($\lesssim 2000$\,km\,s$^{-1}$) Balmer emission lines \citep{Schlegel2006,Filippenko1997,Gal-Yam2017}. Rather than a signature of the explosion itself, this spectral specificity is presumably the result of the interaction between the fast ejecta and a low-velocity, dense, hydrogen-rich, circumstellar medium that surrounded the star before it exploded. During the last decade, the physical picture governing SN IIn explosions and the wider family of ``interacting'' SNe -- SNe whose radiation can be partially or completely accounted for by the ejecta crashing into a dense surrounding medium -- has aroused a lot of interest.

SNe IIn are presumably powered (at least partially) by the conversion of the ejecta kinetic energy into luminosity. This is a broad family of objects, with a wide variety of CSM masses ranging
from $\gtrsim10$\,M$_{\odot}$ (e.g., \citealt{Ofek2014}) to $0.01-0.1$\,M$_{\odot}$\,y$^{-1}$ (e.g., \citealt{Kiewe2012}).
Such low-mass CSM events evolved faster, and are less luminous compared with high-mass CSM events (Ofek et al. 2014b).
At the low-CSM mass, the IIn class is likely related to the flash spectroscopy SN events which have estimated CSM masses of the order of $\sim10^{-3}$\,M$_{\odot}$ (e.g., \citealt{Gal-Yam2014,Yaron2017}), which are confined to the close vicinity of the progenitor star. ZTF will provide a unique insight into these objects.
Type IIn Balmer narrow lines may
persist for days (``flash spectroscopy'', \citealt{Gal-Yam2014,Khazov2016,Yaron2017}, weeks (e.g., SN 1998S, \citealt{Li1998,Fassia2000,Fassia2001}; SN 2005gl, \citealt{Gal-Yam2007}; SN 2010mc \citealt{Ofek2013a}; PTF\,12glz \citealt{Soumagnac2018}), or years (e.g., SN 1988Z, \citealt{Danziger1991,Stathakis1991,Turatto1993,VanDyk1993,Chugai1994,Fabian1996,Aretxaga1999, Williams2002,Schlegel2006,Smith2017}; SN 2010jl \citealt{Patat2011,Stoll2011,Gall2014,Ofek2014}). With its cadence and continuous coverage, ZTF will allow the study of the early and late photometric and spectral properties of interacting SNe and reveal the complete spectral evolution of these events. This will allow the investigation of the various physical scenarios leading to the presence of interaction signatures in the spectrum of SNe.
ZTF has the potential to identify and followup hundreds of Type IIn SN. This will further lead to characterization of their CSM masses (e.g., Kiewe et al. 2012, Ofek et al. 2014c), probing their ejecta shock velocities which is a probe of the internal explosion mechanism (e.g., Ofek et al. 2014c),
and CSM geometry via the spectral and light curve evolution (e.g., Soumagnac et al. 2018).

Furthermore, ZTF will give new insights into the precursors of SNe IIn. 
In recent years it has become clear that a large fraction of SN progenitors show outbursts accompanied by large mass ejections (e.g., $\gtrsim 10^{-3}$\,M$_{\odot}$)
months to years prior to the terminal explosion of the star as a SN. In some cases, there are direct observations of such prior luminous outbursts (e.g., \citealt{Foley2007, Pastorello2007, Mauerhan2012,Prieto2013,Corsi2014,Fraser2013,Ofek2013a, Ofek2013c, Ofek2014b, Ofek2016,Strotjohann2015,Nyholm2017}) and in other cases we detect high excitation emission lines, presumably due to the presence of massive circumstellar material around the SN progenitor (e.g., \cite{Gal-Yam2014,Khazov2016,Yaron2017}). These observations suggest that the final evolution of massive stars is not well understood, as such eruptions do not occur in standard stellar evolution codes. This may be important for a better understanding of the SN explosion mechanism as these final stages determine the initial conditions to explosion simulations (e.g., \citealt{Arnett2011,Quataert2012,Shiode2014,Fuller2017,Fuller2018}). ZTF will cover the locations of multiple SNe IIn multiple times prior to the actual explosions, and will thus allow to better study the precursors of SNe IIn. The public survey, with its all-sky footprint and uniform cadence ($g$ and $r$ band every three nights) will be particularly well suited to perform such study.

\subsection{Superluminous supernovae}
\label{sec:secSLSN}

Superluminous supernovae (SLSNe) are a rare class of transients with peak luminosities 10--100 times higher than ordinary core-collapse and Type Ia SNe, and total radiated energies in excess of $10^{51}$~erg (see e.g., \citealt{gal12} for a review). Their enormous energies cannot be explained by standard supernova models, and their progenitors and energy sources are still debated. Suggestions include either a central energy source (such as magnetar spin-down; \citealt{kb10,woo10}), strong interaction with dense CSM converting kinetic energy to radiation \citep{wbh07,ci11,sbn+16}, or the pair-instability explosion of a very massive star \citep{brs67,gmo+09}. 

ZTF will make progress in SLSN science in multiple ways. First, while more than 100 SLSNe from many different surveys have been reported to date (e.g., \citealt{dgr+17} (PTF), \citealt{lcb+18} (PS1)), fundamental population properties such as the SLSN rate are still only poorly constrained \citep{qya+13,msr+14,pss+17}. ZTF's combination of sky coverage and cadence is ideal for selecting a large sample of SLSNe in a systematic fashion, and thus determining population properties. More generally, increasing the sample of SLSNe with well-determined explosion dates and rise times, as well as color information also on the rise, is important for constraining the progenitors and energy sources of SLSNe, since the rise time encodes information about the diffusion timescale and hence the progenitor mass. For slowly-declining SLSNe, the rise time is the main discriminator between pair-instability and central-engine models (e.g., \citealt{nsj+13, lcb+16}). 

Finally, ZTF's capability to study young SNe (sec.~\ref{sec:InfantSN}) extends to studying the early phases of SLSNe. Some SLSNe show a precursor ``bump'' on the rise, with typical timescales of $\sim 10$~days \citep{lcd+12,nsj+15,ssd+16}. It has been suggested that this feature is ubiquitous \citep{ns16}, but the presence of such a bump is poorly constrained due to a lack of well-sampled early time data in the majority of SLSNe. Understanding the physical nature of this precursor emission, and more generally what it implies for the explosion mechanism and extended structure of the progenitor star, holds great promise in shedding light on these enigmatic explosions. 

\begin{figure}
\includegraphics[width=\columnwidth]{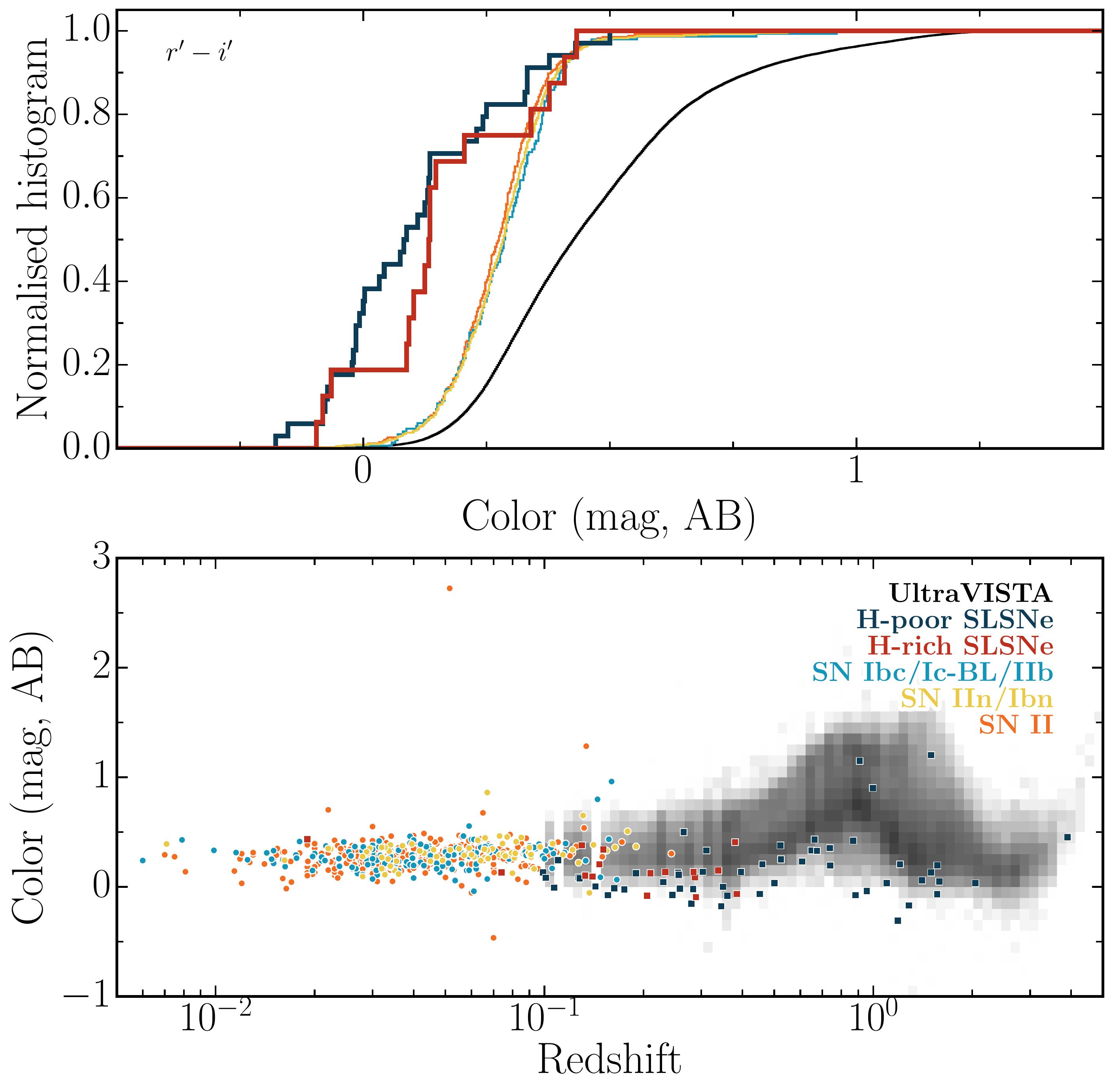}
\caption{$r'-i'$ colour distribution of 500 PTF/iPTF core-collapse SN host galaxies, SLSN host galaxies and galaxies from the UltraVISTA survey. The top panel shows the cumulative distribution for all samples at $z<0.7$. SLSN hosts are found in a part of the parameter space that is sparsely populated by galaxies, in general. The average host of an H-poor/-rich SLSN is 0.2/0.15~mag bluer than that of a regular core-collapse SN. Hence, host galaxy properties could be used to identify infant SLSNe in the ZTF alert stream. Figure adapted from Schulze et al., in prep.}
\label{fig:slsn}
\end{figure}

While the light curve of the precursor can be recovered from ZTF data, securing spectra of the precursor requires detecting SLSNe before their light curve rises for several tens of days and before their luminosity exceeds $-21$ mag. SLSN host galaxies could be very useful in identifying infant SLSNe. SLSNe are preferentially found in star-forming dwarf galaxies with stellar masses of $<10^{10}~M_\odot$ and metallicities of $<0.4$ solar metallicity \citep{Lunnan2014a, Leloudas2015a, Perley2016a, Schulze2018a}, whereas ordinary core-collapse SNe are found in more evolved galaxies \citep[e.g.][]{Kelly2012a}. To illustrate this better, we display in Fig. \ref{fig:slsn} a subsample of 500 core-collapse SN host galaxies from PTF and iPTF surveys with detected hosts in $r'$ and $i'$ band (for details see Schulze et al. in prep.), 53 H-poor and 16 H-rich SLSNe from \citet{Schulze2018a} and galaxies from the UltraVISTA survey \citep{McCracken2012a}. SLSNe are found in a part of the parameter space that is sparsely populated by galaxies in general. Moreover, the average host of H-poor/-rich SLSNe is 0.2/0.15~mag bluer than that of a regular core-collapse supernova (at lower redshift). Hence, host galaxy properties could be a valuable diagnostic to select infant SLSNe in real-time.

\subsection{Exploring the diversity of relativistic explosions}

In the final collapse and explosion of a massive star,
$>$10$^{51}$ erg of kinetic energy are liberated 
as the iron core collapses to a neutron star or a black hole, driving a spherical shock that unbinds the star (a supernova).  A small subset ($\sim$0.1\%) of these explosions exhibit even more extreme behavior: a relativistic bipolar jet is launched, drills through the envelope, and escapes to interstellar space. The jet produces a long-duration gamma-ray burst (GRB) lasting several seconds, and its collision with the circumstellar medium (CSM) produces an ``afterglow'' that radiates across the electromagnetic spectrum for days to months \citep{Rees1992MNRAS}.

To date, $\sim20$ SNe have been spectroscopically confirmed in association with GRBs,
beginning with the coincident discovery of GRB\,980425 and SN\,1998bw at $d = 40\,$Mpc \citep{Galama1998,Kulkarni1998},
All GRB-SNe have had envelopes stripped of hydrogen and helium (Type Ic) and high measured photospheric velocities ($\gtrsim 20,000$\,km s$^{-1}$). These ``broad-lined" (BL) Ic SNe constitute $\sim 1\%$ of the local core-collapse rate,
and their association with GRBs has led to the suggestion that GRBs and at least some Ic-BL SNe arise from a single explosion mechanism \citep{Barnes2017,Sobacchi2017}.

A major focus of scientific investigation over the past 20 years has been to understand the connections between these energetic Ic-BL supernovae with successful, observed jets and ordinary (non-relativistic) SNe without them.  A minimum prerequisite for the launch of a jet is the formation of a ``central engine'' from the collapsing core: a highly magnetized neutron star or rapidly-accreting black hole. But even if such an engine forms, a number of other conditions must also be met for us to observe the jet as a GRB.  First, the jet must be nearly baryon-free (else the available energy is insufficient to accelerate the ejecta to ultra-relativistic velocities), and gamma-ray emission will be stifled by pair-production.  Next, the jet must successfully escape the star without being smothered by the stellar envelope.  Finally, the jet must be directed at Earth.

If any of these conditions are not met, a variety of different empirical phenomena are predicted:

\noindent \textbf{(a)} A jet with too many baryons ($>10^{-4}\,M_\odot$) is known as a \textbf{dirty fireball}.  Given the energy budget of the explosion it can attain only a moderate Lorentz factor ($\Gamma$ $\sim$ 5--10). So, while it successfully escapes the star and should produce a luminous afterglow, it will not produce \emph{significant} high-energy emission and thus not trigger gamma-ray instruments \citep{Dermer1999}. 
 
\noindent \textbf{(b)} A jet which fails to escape the stellar envelope is sometimes termed a \textbf{choked jet} or a failed jet \citep{Meszaros2001}.  However, the jet energy may be transferred into a shock wave that propagates through the star and breaks out at the surface: this may produce a low-luminosity gamma-ray burst (LLGRB) \citep{Bromberg2011,Nakar2015} and a Type Ic-BL SN.
 
\noindent \textbf{(c)} In most cases, the viewing angle exceeds the jet half-opening angle.
Such an \textbf{off-axis jet} (clean or dirty) that escapes the star  will result in an \textbf{orphan afterglow} \citep{Rhoads1997}. The relativistic beaming of such an event means that there will be suppressed (or a lack of) observed gamma-ray emission. The afterglow will brighten as the shock slows and the relativistic beaming cone widens to include Earth (e.g., \citealt{vanEerten2010}).

A census of these phenomena is required to quantify key physical processes in core-collapse SNe and their connection to relativistic transients. How many SNe actually produce central engines?  Why do ``classical" GRB jets accelerate only a tiny fraction of their mass: is the fractional mass fundamental to the phenomenon, or the tip of the iceberg of a wider range of jet phenomena? In particular, do LLGRBs result from  jets getting choked within the star? And, finally, how accurate is our understanding of beaming in GRBs?

We are addressing this area via three surveys. 
The partnership moderate cadence survey (sec. \ref{sec:InfantSN}) is well suited to find 
rapidly fading afterglows. In fact, the moderate cadence survey was based on the success of
a  pilot project undertaken with PTF (which resulted in a cosmological afterglow candidate, PTF11agg; \citealt{Cenko2013}).  
A nightly cadence survey, another boutique survey, is squarely aimed at LLGRB and LLGRB-like SNe such as SN\,2006aj \citep{Soderberg2006} and iPTF\,16asu \citep{Whitesides2017} as well as orphan afterglows. We have undertaken archival analysis of iPTF data and devised excellent
filters to reject false positives for these two surveys (M dwarfs and dwarf novae; \citealt{Ho2018}).
Finally, the public all-sky three-night survey  will result in SEDM classifications of 500 SNe per semester, of which a few percent will be Ic/BL SNe. 

\begin{figure}[htbp]
   \centering
   \includegraphics[width=3in]{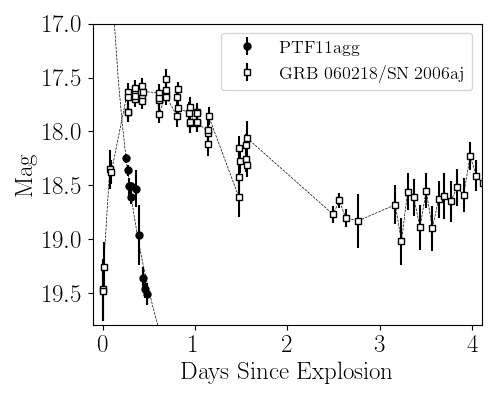} 
   \caption{\small
   Light curves from the candidate on-axis dirty fireball PTF11agg and the low-luminosity GRB\,060218 / Ic-BL SN\,2006aj.
   The rapid decay of PTF11agg is attributed to on-axis fading afterglow. The rise and fade of of SN\,2006aj at early times is likely due to shock cooling emission of SN ejecta.
   }
   \label{fig:PTF11agg}
\end{figure}

\subsection{Rare transients in the local universe}

The luminosity gap between novae and supernovae has recently been populated with a variety of faint and fast evolving transients in the local universe \citep{Kasliwal2012b}.
PTF/iPTF, with its untargeted wide-field search and committed spectral classifications
with the Palomar 200-inch and SEDM (peaking at 700 classifications per year) resulted
in several interesting transients including  rapidly evolving Type Ic SNe like SN 2010X and iPTF 14gqr (\citealt{Kasliwal2010}; \citealt{De2018}; see also SN 2005ek; \citealt{Drout2013}), and significantly increased the samples size of calcium-rich gap transients \citep{Perets2010,Kasliwal2012a}. However, the physical nature of these explosions remains largely debated as their faint and fast evolving light curves (and hence low ejecta mass) point to very low mass progenitors unlike the population of Type Ia and core-collapse SNe. Suggested progenitor channels include tidal disruptions of low mass white dwarfs by a neutron star or black hole \citep{Sell2015}, He shell detonations on the surface of white dwarfs \citep{Bildsten2007} or core-collapse explosions of highly stripped massive stars \citep{Kleiser2014, Tauris2013}. Yet, the small number of confirmed events leaves considerable uncertainty about the intrinsic properties of these intriguing transients.

ZTF, with its order of magnitude improvement in survey speed over PTF, will be a powerful tool to discover large samples of these faint and fast evolving explosions in the local Universe. This will aid not only in understanding the distribution of the intrinsic properties of this population (e.g., ejecta masses, explosion energies and peak luminosities) but also shed light on their progenitors via their host environments. For instance, the old and remote environments of the class of Ca-rich gap transients (a total of eight confirmed events thus far) suggest their association with a very old progenitor population that has traveled far away from their host galaxies, consistent with progenitors arising from old white dwarf binary systems (\citealt{Lunnan2017}; see also \citealt{De2018}). On the other hand, the star forming host galaxies of the fast Type Ic SNe like SN 2010X and SN 2005ek are consistent with white dwarf progenitors as well as core-collapse explosions of ``ultra-stripped'' massive stars \citep{Moriya2017}.

The number of false positives is significantly reduced by the requirement that candidates
be in the apparent proximity of galaxies with $z<0.05$. The resulting spectroscopic
load (for classification) then becomes manageable with the resources available to the
ZTF partnership (e.g.,\ Palomar 200-inch, Nordic Optical Telescope, Liverpool Telescope and
peer-reviewed Gemini allocations).
Early follow-up allows one to directly constrain the pre-explosion properties of the progenitor star. Such techniques have already been demonstrated to be a powerful probe of the nature of the progenitor in the case of Type Ia \citep{Nugent2011} and core-collapse SNe \citep{Yaron17}, and will be important for shedding light on the progenitors of these rare transients. Finally, ZTF will also be important in constraining the rates of these transients (which are otherwise poorly constrained due to the small number of events) that likely have important effects on the chemical evolution of the universe \citep{Mulchaey2014}.
Current estimates of the rate of Ca-rich gap transients that include the survey efficiencies of PTF suggest that their rates are nearly half of the Type Ia supernova rate. This indicates that ZTF is expected to find more than 20 such events per year \citep{Frohmaier2018}. Rate constraints will also benefit from such events initially detected by on-going time domain surveys.

\subsection{A larger sample of stripped envelope supernovae, and their host galaxies}\label{sec:SESN}
The stripped envelope supernovae (SE~SNe) samples currently available in the literature are mainly targeted (i.e., 34 SE SNe in \citealt{Taddia18a}, CSP), and non-homogenous (i.e. collections by \citealt{Cano13}, \citealt{Lyman16}, and \citealt{Prentice16}), or rather small (i.e., 20 SNe in \citealt{Taddia15}, SDSS II).
ZTF will allow us to build a large, homogeneous SE~SN sample. Its untargeted nature will diminish the bias toward metal-rich SN host galaxies, as it will enable finding SE~SNe also in low-luminosity and low-metallicity galaxies. 
With this sample, the main scientific questions are related to understanding the nature of the progenitor stars of SE~SNe. 
By modeling the light curves of our SE~SNe sample we aim to determine the range of the SN explosion parameters. Beside the semi-analytic Arnett model \citep{Arnett82}, we can make use of more sophisticated hydrodynamical codes, such as HYDE \citep{Ergon15} and SNEC \citep{Morozova15}, to estimate the explosion parameters. 
We know from the literature that most SE~SNe have relatively narrow light curves, suggesting moderate to low ejecta mass (order of 2-4 M$_\odot$). However, with iPTF we initiated a sample study of light curves, identifying a number of unusual SE~SNe.
For example, we discovered a dozen SE~SNe with broad light curves that might have massive progenitors, e.g., iPTF15dtg \citep{Taddia16} and PTF11mnb \citep{Taddia18b}, and/or alternative powering mechanisms (e.g., magnetar). We also studied iPTF14gqr, a SN with a much narrower light curve than average, very likely arising from an ultra-stripped progenitor \citealt{De2018}. 
We expect to considerably increase the size of these samples with the forthcoming ZTF data. To increase the sample size is particularly important for these rare SNe, which we have hitherto followed only in a few cases. The stellar Initial Mass Function is steep, and to sample the most massive progenitors simply requires a large number of events.
With ZTF we will be able to discover these SNe routinely, and we aim to follow them up to classify them and to characterize their light curve shapes.

With iPTF we also discovered and investigated the unusual presence of early light curve excesses in some SNe Type Ic. 
This is compatible with the presence of an extended envelope (tens to hundreds of solar radii) around their progenitor stars (e.g., \citealt{Taddia16}). With the higher cadence of ZTF, we aim to routinely study these early emission excesses,
to get a better constraint on the progenitor radius of SE~SNe. 
ZTF will also provide more early color information ($g$ and $r$ band from P48) of the SN emission. This will allow us to build bolometric light curves and temperature profiles at early epochs, which provide information on the degree of $^{56}$Ni mixing in the SN ejecta. 
In summary, with ZTF we aim to observe a substantial sample of SE~SNe with tight pre-explosion limits, pre-maximum coverage and multiband light curves over a range of host galaxy properties.

\subsection{Failed supernova shock breakouts}

Core collapse of a massive star may result in a ``failed'' SN, where the core promptly forms a black hole after the accretion shock fails to explode the star.  While this fate has long been suggested for very high mass stars at low metallicity, there is mounting evidence that failed SNe may also occur in red supergiants (RSGs) with solar metallicity.  First, there is a dearth of $>$$18~M_{\odot}$ RSG SN progenitors---the ``missing RSG problem'' \citep{Kochanek08,Smartt09,Horiuchi14,Smartt15}. 
Also, a significant fraction of core-collapses resulting in failed SNe naturally explains the gap between the neutron star and black hole mass distributions \citep{Kochanek14}.  
The most dramatic evidence is the disappearance of a $10^{5.3}~L_{\odot}$ RSG \citep{Gerke15,Adams17} discovered by an ongoing survey monitoring a million RSGs in nearby galaxies with deep optical imaging \citep{Kochanek08}.  

Searching for disappearing core collapse progenitors is observationally expensive and cannot feasibly be scaled up enough to tightly constrain the rates and progenitor properties.  
Moreover, with this approach candidates are only identified months (or years) after core collapse,  making detailed observations of the event and its immediate aftermath impossible.

However, the disappearance of the progenitor is not the only possible signature of these events. Models predict that even if the energy released by the core collapse of a RSG fails to result in a SN, the loss of gravitational binding energy from the neutrino emission may result in a low-velocity ($\sim$$100~\mathrm{km}~\mathrm{s}^{-1}$) ejection of the weakly-bound hydrogen envelope \citep{Nadezhin80,Lovegrove13}, giving rise to a faint ($\sim$$10^{6}~L_{\odot}$) but long-lived (months -- years) recombination powered transient \citep{Lovegrove13,Fernandez18}.  
Though the temporal sampling is coarse, the observations of the failed SN candidate reveal a several month long transient consistent with this prediction \citep{Adams17}.  Given the likely low rate of failed SNe, this type of transient is too faint to be discovered with supernova surveys, but with the ZTF the shock breakout associated with these events could be discovered for the first time,  triggering spectroscopic follow-up and a search for the subsequent fainter recombination-powered transient.  

Although the shock breakouts of normal SNe are very short (seconds to hours) and radiate primarily in the UV and X-rays, the shock breakout from the low-energy, neutrino-mediated shock of a failed SN is predicted to have a duration of a few days and be thermalized to a temperature of $\sim$$10^{4}$ K, with a luminosity of $\sim$$10^{7}~L_{\odot}$ \citep{Piro13,Lovegrove17,Fernandez18}. 

Observations of both the shock breakout and the subsequent recombination phases would provide a unique confirmation that a failed SN has occurred.  The shock breakout luminosity, temperature, and duration together with the luminosity and duration of the subsequent recombination powered transient can constrain the progenitor radius, the explosion energy, and the ejected mass.

Though the expected ZTF discovery rate of failed SN shock breakouts is low, this approach represents the only feasible way to promptly discover and observe the birth of a new black hole from stellar core-collapse for the very first time. The only false positives are novae (which, due to their luminosity function, are limited to galaxies no further than 10\,Mpc). ZTF is well suited to this project given its depth.

\subsection{Bright Transient Survey}


Two science drivers motivated the Bright Transient Survey (BTS).
First, one of the approaches for finding electromagnetic (EM) counterparts to 
neutron star mergers is to target galaxies in the localization constraints
(including redshift) provided by the gravitational wave (GW) facilities
(e.g., \citealt{Gehrels16}). However, the quantitative efficiency of this method requires the knowledge of the redshift completeness fraction (RCF) of these
catalogs. We measure RCF using SNe as markers of galaxies (regardless of their
luminosity).
Preliminary estimates of the RCF find that $\sim$75\% of $z < 0.03$ galaxies are cataloged, based on observations of $m_\mathrm{peak} < 17\,\mathrm{mag}$ SNe from the ASAS-SN survey \citep{Kulkarni18}.
Next, there is widespread recognition of increased precision for
Ia SN cosmology at low redshift, $z<0.15$ (and discussed in great
detail in sec.\ \ref{sec:SNIa}).  To satisfy these two projects
a significant fraction of SEDM has been set aside to spectrally classify 
bright transients ($\lesssim 19$\,mag) with the expectation of completeness to
$18.5$\,mag. Such bright transients will not only be found by the ZTF public survey but also by ASAS-SN, ATLAS and PS. We plan to publish a yearly catalog spelling in detail observational conditions so that the sample can be used to compute reliable rates. With the data in hand it appears that we are on course to classify 500 bright transients every semester.

\section{Multi-messenger Astrophysics}
Multi-messenger astrophysics is a growing methodology in astronomy and to this end
we have built-in Target-of-Opportunity (TOO) capability.
Multi-messenger astrophysics has three science objectives: (i) identifying electromagnetic (EM) counterparts to neutrino triggers from IceCube; (ii) identifying afterglows to short hard gamma-ray bursts from the {\it Fermi} satellite; and (iii) identifying electromagnetic counterparts to gravitational wave (GW) triggers from LIGO/Virgo. 

\subsection{Identifying electromagnetic counterparts to neutrinos}


The IceCube Neutrino Observatory has discovered a diffuse flux of high-energy neutrinos~\citep{Aartsen:2015knd, Aartsen:2013jdh}. However, until recently no compelling evidence for spatial or temporal clustering of events had been identified and the origin of the neutrinos was unknown~\citep{Aartsen:2016oji,2015ApJ...807...46A}. The consistency of the spatial distribution with an isotropic flux points to a predominantly extragalactic origin for the neutrinos. Multi-messenger studies are key to identifying the neutrino sources, through detection of their EM counterparts. ZTF's all-sky coverage and high cadence will play a crucial role in detecting potential optical counterparts to astrophysical neutrinos, such as flaring blazars, choked-jet supernovae~\citep{Senno:2015tsn}, CSM-interacting SNe~\citep{Murase:2010cu, Zirakashvili:2015mua}, and tidal disruption events~\citep{2017PhRvD..95l3001L}. Our goal is to identify the neutrino sources through two complementary approaches:


Firstly, a target of opportunity (ToO) program will select the most promising astrophysical neutrino candidates in real-time~\citep{Aartsen:2016lmt} from IceCube, and trigger rapid follow-up observations with ZTF to target fast-evolving transients (such as GRB afterglows). ZTF's ToO marshal will enable prompt collection of early photometry. These observations, in combination with upper limits provided by the regular all-sky survey, will allow tight constraints to be placed on the explosion time of GRBs or choked-jet SNe. Such temporal constraints are crucial to establish the causal connection between the neutrino and the potential optical counterpart. The effectiveness of ToO follow-up in identifying possible optical counterparts is already well-established.

One interesting candidate, SN PS16cgx, was found during the Pan-STARRS optical follow-up of the first publicly released high-energy neutrino alert \citep{2016GCN..19381...1S}. With a tentative classification as a broad-lined Type Ic, the object could belong to the rare class of objects that is also associated with long GRBs, and hence a potential neutrino source. 

Another follow-up of a more recent high-energy neutrino event revealed a spatially coincident blazar, TXS\,0506+056, which was found by {\it Fermi}-LAT to be in flaring state~\citep{IceCube:2018dnn}. The coincidence triggered further multi-wavelength follow-up, leading to the discovery of very-high-energy gamma-ray emission by MAGIC. Study of archival optical data revealed a rise of $\sim0.5$ mag in V-band over the preceding 50 days. Those findings are consistent with Fermi blazars contributing $<10\%$ to the diffuse neutrino flux~\citep{2018arXiv180704748M}.


Secondly, while only a handful of the highest-energy neutrinos, with a $>50\%$ chance to be of astrophysical origin ($\sim 10$ per year), are suitable for the ToO program, there are many more detected neutrinos that could also have optical counterparts. We can utilize ZTF's all-sky survey to access these lower-energy cosmic neutrinos, which are buried in a background of atmospheric neutrinos. With an all-sky real-time search, in which we correlate all optical transients found by ZTF with all neutrino candidates detected by IceCube, we will target potential optical transient counterparts (e.g. SNe, TDEs) accounting for position, time and neutrino energy. In particular, an online stream of approximately $100$ neutrinos per day will be cross-matched with all detected ZTF transients during each night of observation. Positive correlations will trigger a dedicated follow up campaign for potential optical transient counterparts which will also enable us to acquire a
complete flux-limited catalog (to 20th mag) of classified sources as potential neutrino counterparts. IceCube's most sensitive sky region [the Northern sky]~\citep{Aartsen:2016oji} is excellently matched by ZTF's coverage of the Northern sky.

The discovery of the origin of high-energy neutrinos would be a breakthrough for the emerging field of neutrino astronomy, and would furthermore reveal the much sought-after sources of high-energy cosmic rays. 
More specifically, the detection of neutrinos from choked-jet SNe would offer a direct window to the internal dynamics of those sources. It would constrain the composition, energetics and Lorentz boost
factor of relativistic outflows leaving the collapsing star, and resolve the currently
uncertain emission mechanism for GRBs. Deciphering the processes in the cores of collapsing stars hidden from electromagnetic observations is one of neutrino astronomy's key science goals.

\subsection{Identifying afterglows to short hard gamma-ray bursts}
The recent discovery of broadband electromagnetic radiation associated with gravitational waves from a binary neutron star merger \citep{Abbott+2017a} has ushered in a new era of multi-messenger astrophysics.  One of the more unexpected results from this discovery was the detection of a low-luminosity short gamma-ray burst (GRB\,170817A; \citealt{Goldstein+2017}) just 1.7\,s after the binary neutron star (BNS) merger.  With $E_{\gamma,\mathrm{iso}} \approx 3 \times 10^{46}$\,erg \citep{Abbott+2017b}, GRB\,170817A is 3--4 orders of magnitude less energetic than all previous short GRBs with secure redshift measurements.

The explanation for this low-luminosity gamma-ray emission remains hotly debated.  One possibility is that an ultra-relativistic jet was launched following the binary neutron star merger, but our viewing angle is slightly off-axis (though still within the envelope) of a ``structured jet'' \citep{Abbott+2017b}.  As a result, the gamma-ray luminosity we observe is significantly reduced, but some other observer in the Universe would have seen a classical short GRB following the binary neutron star merger.  Alternatively, the gamma-ray emission may have been powered by a (quasi-)spherical, mildly relativistic outflow.  Such emission may arise naturally from the ``cocoon'' formed when a jet fails to penetrate the neutron-rich material dynamically ejected prior to the merger \citep{Kasliwal+2017,Gottlieb+2017}.

Regardless of the origin, it is clear that the gamma-ray emission from a NS merger can be observed from outside the narrow opening angle of the ultra-relativistic jets of classical (i.e., high $E_{\gamma,\mathrm{iso}}$) short GRBs.  Thus, low-luminosity short GRBs may offer a new means to identify the r-process kilonovae following neutron star mergers, independent of any gravitational wave trigger.

No short-duration GRBs have been conclusively identified within 200\,Mpc (the horizon distance for BNS mergers from Advanced LIGO at design sensitivity) to date, despite dozens of robust host associations from the well-localized \textit{Swift} sample (e.g., \citealt{Berger2014}).  But the Gamma-Ray Burst Monitor (GBM; \citealt{Meegan+2009}) on-board the \textit{Fermi} satellite triggers on $\approx 4 \times$ more short-duration GRBs per year than \textit{Swift} (with even more detected via ground-based pipelines; \citealt{Briggs+2017}).  Few (if any) of these GBM short-duration GRBs are followed up with optical facilities \citep[see e.g.,][]{Golkhou2018ApJ}, due primarily to their coarse localizations from several hundred up to $\sim 1000$\,deg$^{2}$.

With the large field-of-view and automated transient identification pipeline of ZTF, we will follow a sample of short-duration GRBs from the \textit{Fermi}-GBM to search for kilonova counterparts.  While most such GRBs will be at distances $\gg 100$\,Mpc (the approximate distance out to which the GBM could detect GRB\,170817A; \citealt{Abbott+2017b}), within this volume the rate of BNS mergers is $\approx 6$\,yr$^{-1}$ \citep{Abbott+2017a}.  If all BNS mergers have a $\gamma$-ray signal of comparable luminosity to GRB\,170817A (as may be expected in cocoon models), ZTF will be capable of finding several counterparts per year, even before the next LIGO and Virgo observing run begins.  For those more distant events, ZTF will be sensitive to the bright but rapidly fading afterglow emission, allowing robust host association and redshift and offset measurements.

\subsection{Identifying electromagnetic counterparts to gravitational wave transients}
Pinpointing EM counterparts to neutron star mergers has the potential to unlock a wide range of new astrophysics, as illustrated by GW170817.  For instance, detailed photometry and spectroscopy coupled with reliable rate estimates 
will quantify how prolific a site of r-process nucleosynthesis they are and whether they can explain the observed Solar abundance of heavy elements \citep[e.g.,][]{Drout2017,Kasliwal+2017,Kasen+2017,2018arXiv180101141H}.  EM counterparts are crucial to reliable measurements of the Hubble constant \citep{Schutz86,2017Natur.551...85A} which is still a topic of interest \citep{2018arXiv180410655R}. Neutron star mergers are also unique laboratories to study jet physics \citep[e.g.,][]{2018MNRAS.tmp.1056L,2017ApJ...850L..24G,2017arXiv171203237L,2018arXiv180305892G,2018MNRAS.tmp.1009N}
especially the wide-angle mildly relativistic cocoon breakout seen in GW170817 \citep{Hallinan+18,Mooley+18,Dobie+18,2018arXiv180502870A,Troja+18}.

With its  combination of mapping speed and depth, ZTF is well poised to identify EM counterparts given its location at Palomar Observatory (facilitating prompt response), although we note that there are other facilities which may be more optimal for this purpose. Based on pessimistic models for optical emission \citet{Ghosh+17} worked to optimize the followup strategy, distributing observations to cover the gravitational-wave error region (which could be $1000\,\,{\rm deg}^2$, depending on the number of detectors involved; \citealt{Singer+14}) and showed that ZTF should be able to detect a significant fraction of sources in the upcoming third GW observing run.  Given how bright the early-time emission was from GW170817 \citep[e.g.,][]{Drout2017,Kasliwal+2017,Arcavi+17,Nicholl+17} we may even be able to see a significant fraction with shorter observations, but we are baselining our plans to 
be able to detect counterparts $10\times$ fainter than GW170817 at 120\,Mpc.

\section{Cosmological distances from Type Ia supernovae}\label{sec:SNIa}
\begin{figure}
\includegraphics[width=0.45\textwidth]{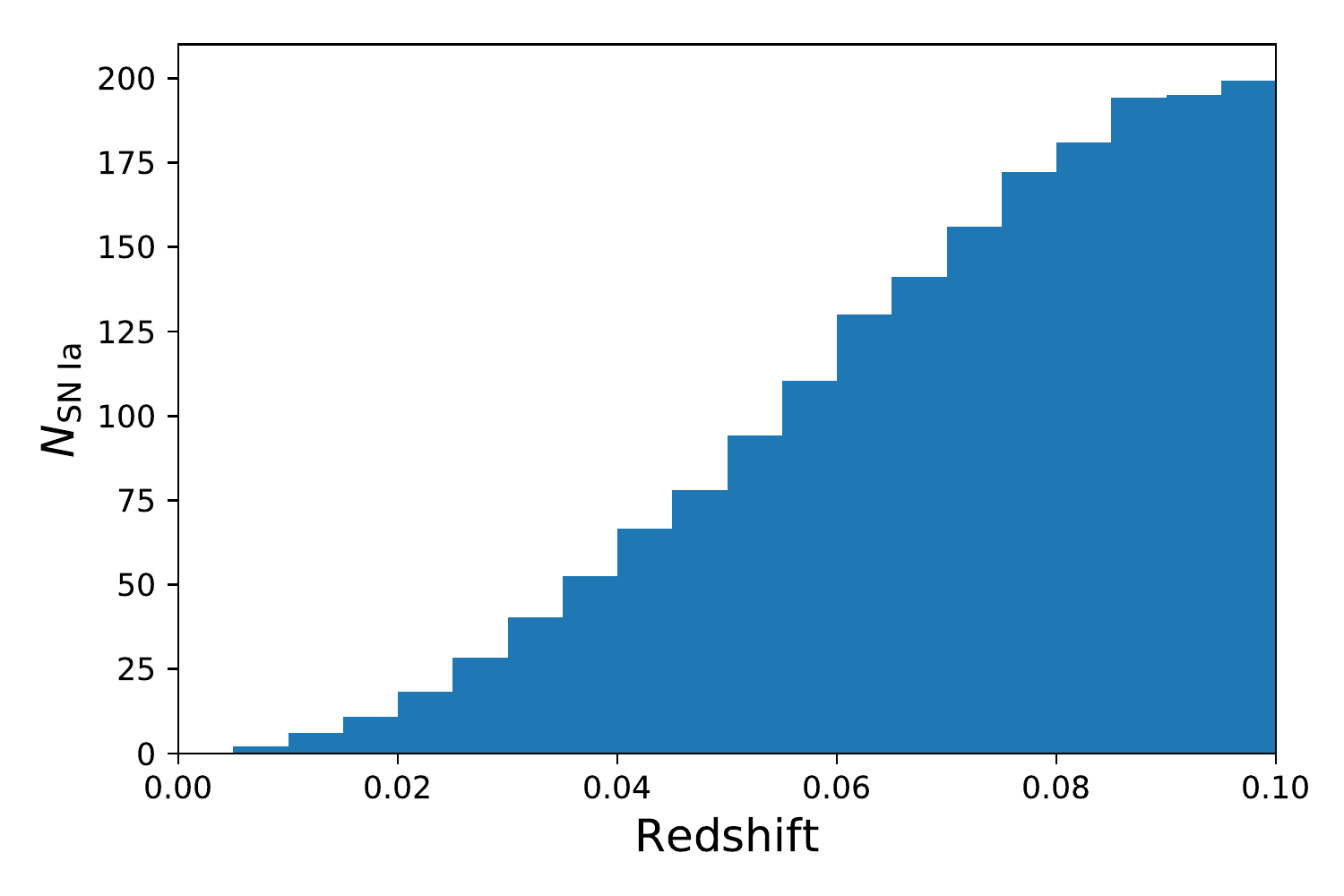}
\caption{Redshift distribution of the 2000 expected SN~Ia cosmology spectroscopic sample, where the upper redshift limit $z=0.1$ is chosen to mitigate the impact from Malmquist bias. Only supernovae discovered >10 days prior to lightcurve maximum are included.}
\label{fig:SNIaz}
\end{figure}

The use of Type Ia supernovae as distance indicators led to the
discovery of the accelerating
Universe \citep{1998AJ....116.1009R,1999ApJ...517..565P}, attributed to
the existence of a new cosmic component dubbed ``dark energy''
\citep[see][for a review]{2011ARNPS..61..251G}. Perhaps the ``simplest'' explanation for dark energy is the one introduced already by Einstein, the 
  cosmological constant, $\Lambda$. Whereas $\Lambda$ would seem to
correspond conceptually to the vacuum energy density expected from
quantum field theory, the measured value of $\rho_{\rm DE}$ is at
least 60 orders of magnitude too small, rendering the association
extremely uncertain. Given the current lack of theoretical
understanding, the quantity used to parameterize the nature of dark
energy (DE) is the dimensionless equation of state parameter, built by
the ratio of the pressure to energy density of the dark energy
cosmological fluid, $w= p_{\rm DE}/\rho_{\rm DE}c^2$, which for the
vacuum energy associated with $\Lambda$ becomes $w=-1$. Using the most
recent SN~Ia compilations in \cite{2014A&A...568A..22B}
and \cite{2017arXiv171000845S} in combination with CMB and BAO data,
attempts have been made to explore alternative dark energy
models. While Einstein's $\Lambda$ in a flat Universe is favored by
the current observations, several competing models based on
well-founded physics remain
unchallenged \citep{2017JCAP...07..040D}. It has since long been
recognised that the low redshift anchoring SN~Ia sample is crucial to
discern between dark energy
models \citep{2001A&A...380....6G,2014A&A...572A..80A}. Besides the
limited statistics, the diverse origin, filter sets and lack of
precise calibration makes the current low-$z$ SN~Ia sample the main
contributor to the systematic uncertainties of the estimates of the
dark energy equation of state \citep{2017arXiv171000845S,
2018MNRAS.475..193F}. The ZTF public survey, in combination with a
partnership $i$-band four-day cadence of 6700 deg$^2$ is expected to
yield nearly 2000 spectroscopically identified SNe~Ia (Feindt et al
{\em in prep.}) over three years, with a sub-percent absolute
calibration with a redshift distribution shown in Fig.~\ref{fig:SNIaz}
and median peak magnitudes of 18.26~mag and 18.32~mag in g- and r-band
respectively.  The ZTF survey can provide a complete and unbiased
SN~Ia sample for $z < 0.1$.

With the ZTF plans of spectroscopic follow-up and management of the
follow-up sample, this should be a multi-band, well sampled
spectroscopically confirmed, un-biased and close to complete data set,
with well understood selection properties.  The data set will be ideal
for studying the supernova population, for example, exploring the
effect of the local host environment on standardization of supernovae
that have been found in other studies \citep[see][and references
therein]{2017arXiv170607697R}. 

ZTF, along with other surveys discovering high rates of SNe~Ia, such as ATLAS, PanSTARRS and ASAS-SN, or the Foundation effort \citep{2018MNRAS.475..193F}, aiming at building up multi-color 
lightcurves for many hundred SNe Ia, are providing the critical anchoring samples for cosmology. The ZTF survey offers specific advantages related to the
understanding of systematic effects that need to be controlled  to make major progress in precision cosmology with SNe~Ia, e.g. corrections for selection effects like Malmquist bias.
These require a knowledge of the underlying supernova population
(rates, luminosity functions, and galaxy occupation distributions)
conditioned on \citep[possibly local;][]{2018arXiv180603849R} host galaxy properties. The difficulties related to the sampling of SNe from different host galaxy populations at low-$z$ compared to the high-$z$ counterparts can be appreciated in Fig.~2 in \citet{2018arXiv181109286J}, showing significant differences between the low and high redshift host galaxy stellar mass. 
Thanks to the better depth than ,e.g., ASAS-SN and ATLAS, the ZTF SNIa lighcurves are sampled over longer time, both before and after peak, and thus better suited for detailed comparisons with high-$z$ samples from LSST and WFIRST, a critical aspect of checks for possible demographic changes in the populations of SNe used for distance measurements. 
Furthermore, the extremely early SN detection, averaging at 13 days
prior to lightcurve maximum, can be used to study the evolution of
color excess, and thereby constrain the location of dust clouds
dimming the SN light, a crucial aspect in the understanding of the
color corrections needed to standardize SNe~Ia for cosmological
distance estimations \citep{2018MNRAS.473.1918B,2018arXiv180309749B}. The ZTF SN Ia sample will also shed light into
the impact of dimming by dust in the intergalactic
medium \citep{Goobar:2018smm}, an effect currently not included in the
cosmological fits with SNe Ia.  ZTF will thus provide an excellent
anchoring sample with which to quantify the key systematic
uncertainties that will limit future high-$z$ surveys from LSST and
WFIRST. Similarly, ZTF will provide an independent SN Ia sample to
measure the Hubble constant, H$_0$, where a nearly 4$\sigma$ tension
has been claimed between the local measurements of the expansion rate
based on SNe~Ia calibrated with Cepheids
\citep{2016ApJ...826...56R,2018ApJ...855..136R,2018arXiv180410655R} and the value derived from 
{\em Planck} measurements of the angle subtended by the sound horizon
as observed in CMB temperature fluctuations
\citep{2016A&A...594A..13P}. Finally, thanks to the large effort on spectroscopic follow-up of both Type Ia and core collapse
supernovae, these data will serve as training samples for
classification of future and ongoing surveys continuing to hold legacy
value for future deeper surveys, even after the completion of ZTF.

The measured redshift of galaxies is given by the combined effects of
cosmological expansion and the peculiar motion induced by the
surrounding gravitational potential. The Type Ia supernovae detected
by ZTF fall into a key distance range where the standardized
luminosity can be used to determine the cosmological distance without
significant dependence on cosmological parameters, while the volume of
the universe is sufficient to produce large samples of supernovae
every year. Such a large sample can then be used to constrain the
correlations between the peculiar velocities in order to study the
structure of the local universe. To first order, this can be done by
measuring the large mode of correlation, a velocity dipole or bulk
flow, which can test whether the nearby distribution of galaxies and
clusters matches our expectation
(e.g.,\ \citealt{2013A&A...560A..90F}). Additionally, the correlation
across all scales can also be used to directly measure the local
growth factor of structure more precisely than has been done before
(e.g.,\ \citealt{2017MNRAS.471.3135H}). This will directly test recent
claims of deviations between the measured structure of the nearby
Universe and the $\Lambda$CDM predictions derived from the {\em
Planck} CMB map \citep{2017MNRAS.465.1454H, 2017MNRAS.471.4412K}. The
ZTF SN Ia peculiar velocity sample will populate the northern
hemisphere in a way that current and future \citep[e.g., TAIPAN and
WALLABY,][]{2017PASA...34...47D} southern galaxy peculiar velocity
samples do not. The much more precise distances derived from SNe mean
that the few thousand ZTF SNe would provide similar statistical
constraints as the many times larger galaxy
samples \citep{2014MNRAS.445.4267K,2017ApJ...847..128H}. More
importantly, the small SN Ia intrinsic dispersion fraction of distance
estimate errors means that the potential for systematic uncertainties
is much reduced \citep{2013MNRAS.432L..90B, 2013A&A...560A..66R,
2018arXiv180101834N}. A further key advantage of the ZTF SN peculiar
velocity sample will be the small and well understood Malmquist bias.

\section{Cosmology with  gravitationally lensed supernovae}

One of the foundations of Einstein's theory of General Relativity is that matter curves the
surrounding spacetime. For the rare cases of nearly perfect alignment between an astronomical
source, an intervening massive object, and the observer, multiple images of a single source can
be seen by the observer, a phenomenon known as strong gravitational lensing.
Gravitationally lensed supernovae (gLSNe), and in particular lensed SNe~Ia, have the potential to directly constrain the expansion rate of the universe through the time delay between images \citep{1964MNRAS.128..307R} due to their well-known light curve shapes. 
Time-delay measurements can also
provide powerful leverage for the studies of dark energy in complementary ways to those from
standard supernova cosmology, BAO, CMB and weak lensing \citep{2011PhRvD..84l3529L}.
Although many strongly lensed galaxies and quasars have been detected to date, finding this special configuration for supernovae has proved extremely difficult: only two multiply-imaged supernovae have been discovered to date \citep{2017Sci...356..291G, 2016ApJ...831..205K}. 

Recently we have overcome these discovery challenges through a novel method to discern gLSNe~Ia in wide-field optical surveys \citep{2017ApJ...834L...5G}. We consider the strong gravitational lensing of SNe~Ia by quiescent (E/S0) galaxies, which have three properties that are useful to identify strongly lensed SNe~Ia. First, normal SNe~Ia are the brightest type of supernovae that have ever been observed to occur in quiescent galaxies. Second, the absolute magnitudes of normal SNe~Ia in quiescent galaxies are remarkably homogeneous, even without correcting for their colors or light curve shapes. Finally, due to the sharp 4000 angstrom break in their spectra, quiescent galaxies tend to provide accurate photometric redshifts from large-scale multi-color galaxy surveys such as the Sloan Digital Sky Survey, DECaLS and in the future LSST. A high-cadence, wide-field imaging survey can leverage these facts to systematically search for strongly lensed SNe~Ia in the following way. Given the photometric redshift, compute the absolute brightness of the SN in the quiescent galaxy and if it is brighter than the normal population of SNe~Ia which should occur there, it is likely a background SN Ia being lensed by the quiescent galaxy. This method has been refined even further to include not only the brightness of the supernova, but the shape of the light curve given the photometric redshift \citep{2018ApJ...855...22G}.

An important property of this search technique is that it does not require the ability to resolve the lensed images to perform discovery. Once lensed SN Ia candidates are identified, they can be confirmed using high-resolution imaging, e.g., Laser Guide Star Adaptive Optics or
space-based imaging such as {\em HST} or, in the future, by the James Webb Space Telescope (\textit{JWST})
and the Wide Field Infrared Space Telescope (\textit{WFIRST}).

\begin{figure}
\centering
\includegraphics[width=1\columnwidth]{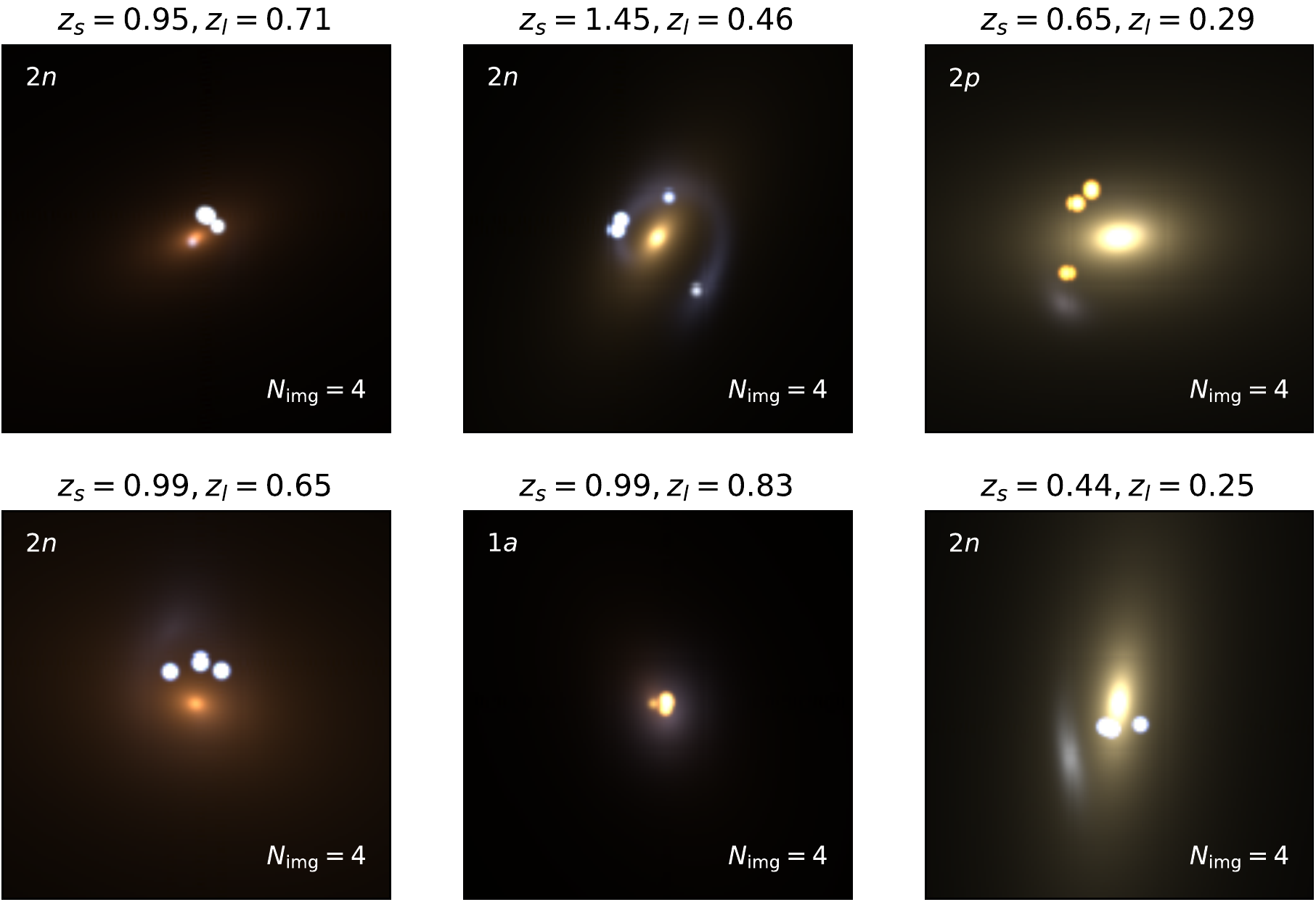}
\caption{Noiseless  $6^{\prime\prime}\times6^{\prime\prime}$ composite $gri$ images of 6 simulated gLSNe, their lens galaxies, and their lensed host galaxies, ``detected'' by ZTF in the simulations of \cite{2018arXiv180910147G}.
 Each image is ``taken'' exactly one night after the transient is detected as a gLSN candidate based on a light curve fit to the simulated ZTF data.
 The FWHM of the seeing on the images is $0.1^{\prime\prime}$, and the pixel scale is $0.04^{\prime\prime}$, identical to that of the UVIS channel of the Wide Field Camera 3 (WFC3) on \textit{HST}. 
}
\label{fig:glsne}
\end{figure}

Given the nominal ZTF survey design coupled with stacking the proprietary data over a ten-day baseline, the detailed Monte Carlo simulations of  \cite{2018arXiv180910147G} show that we would find $\sim$8.6 gLSNe (of all types) per year, of which approximately 1.2 are Type Ia, 2.8 are Type IIP, 0.3 are Type IIL, 0.4 are Type Ib/c, 0.2 are SN 1991T-like, and at least 3.8 are Type IIn, consistent with the calculations of \cite{2011PASP..123...58T} for gLSNe Ia in the ATLAS survey. 
Some examples of these simulated systems are shown in Figure \ref{fig:glsne}.
These lens systems comprise  both doubles and quad systems (like iPTF16geu; \citealt{2017Sci...356..291G}) in a ratio of 2:1 due to the nature of the discovery mechanism. 
The discovered gLSNe have a median $z_s=0.8$, $z_l=0.35$, $\mu_\mathrm{tot}=30$, $\Delta t_\mathrm{max}= 10$ days, $\min(\theta)= 0.25^{\prime\prime}$, and $N_\mathrm{img} = 4$.

\section{AGN and TDEs}
Whilst searches for supernovae and similar explosive phenomena tend to avoid the cores of resolvable galaxies and nuclear-dominated sources, these are the sites of a variety of astrophysical phenomena that relate to the physics of accretion disks and interactions with (super)massive black holes.

\subsection{Tidal disruption events}

A class of transients associated with the nuclei of galaxies is tidal disruption events.  A TDE occurs when a star wanders close enough to a central massive black hole (MBH) to be shredded apart by tidal forces \citep{Lidskii1979, Rees1988}.  A luminous flare is observable when this distance of approach, the tidal disruption radius, is outside the event horizon of the MBH.  These events are rare, with a volumetric rate a factor of 100 smaller than for SNe, with a per galaxy event rate of only $10^{-4}$ yr$^{-1}$ \citep{Velzen2018}.  The rise-time of a TDE is of particular importance, since it scales as $M_{\rm BH}^{1/2}$ \citep{Lodato2009, Guillochon2013}, and can be used as a probe for dormant MBHs otherwise unobservable in distant galaxies.  However, there are only about a dozen TDEs with well-sampled light curves in the literature (see review by \citep{Hung2017}), and only a few with pre-peak light curves.  After peak, the bolometric luminosity is expected to follow the fallback rate of the bound stellar debris, which declines as a t$^{-5/3}$ power-law \citep{Rees1988, Phinney1989,Evans1989}.

\begin{figure}
\vspace{0cm}
\hspace*{-1cm}\includegraphics[width=12.5cm]{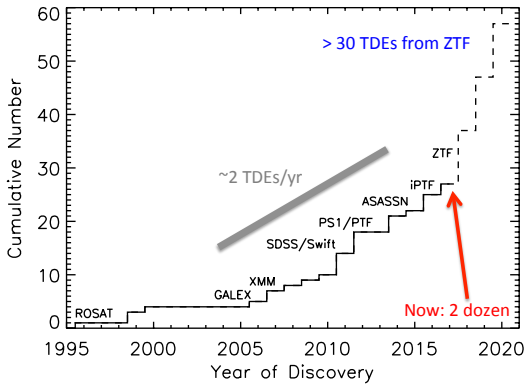}
\vspace{-2cm}
\caption{Cumulative discovery rate of tidal disruption events as a function of time, with the onset of new surveys labeled.  Note the dramatic predicted jump in discovery rate from $\sim 2$ TDEs per year, to $\sim 10$ bright, early TDE discoveries by ZTF with SEDM spectroscopic classification per year.}
\label{FigTDEs}
\end{figure}

From a systematic study of nuclear transients from iPTF, we expect a yield of $4^{+5}_{-3}$ TDEs per month in the ZTF public Northern Sky Survey \citep{Hung2018} (see Figure \ref{FigTDEs}).  Of these TDEs, we expect $\sim 20\%$ to have peak magnitudes $< 19$ mag, bright enough for classification with SEDM, and discovered on the rise to peak.  The selection of TDE candidates will greatly benefit from the $g-r$ color measured by ZTF with $g$ and $r$ observations on the same night, and the measurement of the relative offset of the transient to the host galaxy in the reference image.  TDEs are bluer and have less color evolution than SNe \citep{Velzen2011, Hung2018}, and the majority of AGN can be removed from catalog matches and previous variability history.  Spectroscopic follow-up and UV and X-ray follow-up imaging will be used for classification purposes.  TDEs are characterized by broad helium and/or hydrogen emission lines, a blue, UV-bright continuum, and soft X-ray emission.  A large sample of well-sampled $g$ and $r$ TDE light curves from ZTF, in particular those with a pre-peak discovery, will be critical for mapping the properties of the TDEs to their host galaxy properties and central black hole demographics.  

The rate of TDEs depends on the rate at which stars are scattered into the ``loss cone'' of the MBH, the region of phase space for which a star's orbit passes within the tidal disruption radius, and is a sensitive probe of the nuclear stellar structure of galaxies \citep{StonevanVelzen2016}.  The TDE rate can also be an important probe of MBH demographics, with a potential dependence on the mass of the MBH \citep{Wang2004}, the presence of a binary MBH \citep{Chen2011} or recoiling coalesced MBH \citep{Stone2011}, and the MBH occupation fraction \citep{StoneMetzger2016}. With a statistically significant sample of TDEs from ZTF, we can measure the rates of TDEs as a function of black hole mass and host galaxy type, and look for these theoretically predicted dependencies.  

\subsection{Active galactic nuclei}

Variability is a ubiquitous property of unobscured active galactic nuclei.  In particular, \citet{Sesar2007} showed that $> 90\%$ of type 1 quasars showed optical variability above a level of 2\% in the 290 deg$^{2}$ SDSS Stripe 82 survey on a timescale of several years.  Scaling up to the area of the ZTF public survey, which has a comparable depth of $r \sim 20.5$ mag, we should detect $\sim$ half a million variable AGN. About 1 in 10,000 of these fall into the category of extreme variable AGN \citep{Graham2017} showing significant flaring activity over months to years or other distinct patterns of variability. These may be related to stellar phenomena in the accretion disk or gravitational microlensing. 

With its large survey volume, ZTF will also have the capability to catch AGN in the act of ``changing look'' from a narrow-line (type 2) to a broad-line (type 1) spectrum \citep{Shappee2014, LaMassa2015} and vice versa.  With iPTF, we were able to use the detection of a nuclear transient in an SDSS LINER galaxy to trigger follow-up optical spectroscopy and \textsl{Swift} UV and X-ray imaging to reveal that the galaxy had transformed into a broad-line quasar in $< 1$ yr \citep{Gezari2017}.  The rate of changing-look AGN (CLAGN) is not yet well constrained; however, from a pilot study of spectroscopic follow-up of nuclear transients in iPTF from type 2 AGN, we expect $\sim 10$ per year with ZTF.  Constraining the turn-on/turnoff timescale for CLAGN is important for constraining the mechanism responsible for their spectral transformation. 

One of the challenges in detecting these types of events is that the detection timescale from difference imaging can be several months into the phenomenon -- the day-to-day variability is gradual and so it takes quite some time before the change has become significant enough relative to a reference image to be detected. Fortunately, decade baseline archives of AGN variability over most of the sky are now available, e.g., CRTS \citep{Drake2009}, PTF \citep{Law:09:PTFOverview}, and ATLAS \citep{Tonry2018}, so that the historical behavior of the sources ZTF will see can be characterized and modeled, although the optimal way to combine data from multiple surveys to maximize the information content still needs to be determined. The expected variability can then be predicted, either for a given night or over a particular timeframe, and compared with what is observed. In this way, significant changes from the forecast variability can be identified more quickly and earlier follow-up of the activity, be it changing-look, flaring or something else, triggered. Our models suggest that we can track $\sim 50$ AGN per year in this manner, giving valuable insights into accretion disk mechanics.

\section{Stellar variability}
Variable stars show up in very different flavors ranging from ultra-short period objects like pulsating white dwarfs or ultracompact binaries with periods as short as minutes up to objects with periods of months to years like Cepheid or Mira variables. Over the last two decades many surveys have increased our knowledge in variable stars significantly. These surveys include the Optical Gravitational Lensing Experiment (OGLE; e.g. \citealt{sos15}), CRTS \citep{Drake2009}, PTF \citep{Law:09:PTFOverview}, the Vista Variables in the Via Lactea (VVV; \citealt{saito12}), ASAS-SN, \citep{Shappee2014}) and most recently ATLAS \citep{2018arXiv180402132H}.

 In the transient sky we expect phenomena including: outbursts of young stellar objects, M-star flares, and Nova/dwarf nova eruptions. Archival light curves will allow us to study pulsating and rotating stars as well as compact binaries. Down to a limiting magnitude of 20.5 -- 21 with a median FWHM of $\approx2$arcsec, ZTF will provide one of the best data sets for time-domain astronomy in the Northern hemisphere at low Galactic latitudes, with a median of about $\approx$150 epochs per year at a cadence of minutes to days.

The paradigm of star formation now explicitly includes the concept of episodic accretion. Stars are thought to accumulate some fraction of their mass in the initial spherical infall stage, some fraction during early-stage disk accretion that is punctuated by periods of elevated accretion, and finally the last remaining few to 10\% of their final mass during the optically visible stage of pre-main sequence evolution, which is characterized by mostly low disk accretion rates but also by infrequent bursts. Among the bursts, the most extreme type, called FU Ori events, last decades to perhaps centuries, and involve a thermal or a (gravo-) magneto-rotational instability in the inner $\sim$1 AU of the disk. Bursts with smaller amplitude and shorter duration (months to year-long), called EX Lup type events, may be related to instabilities associated with the interaction region between the disk and the stellar magnetosphere. Outside of the bursts, during routine low-state accretion phases, young star photometric variability occurs with amplitudes between about 2-20\% and on time scales of  1-2 days, with quite diverse light curve shapes. Recent space-based work with CoRoT, MOST, and K2 have illuminated heterogeneity, but also the patterns, characterizing the low-state accretion in young stars.  However, the discovery and study of the more rare EX Lup and FU Ori events, including secure determination of their occurrence rates, remains the domain of wide-field, moderate-cadence, long duration photometric surveys like ZTF.

(Ultra)compact binaries are a rare class of binary systems with periods below a few hours (detached or semi-detached), consisting of at least one compact object. The study of (ultra)compact binaries is important to our understanding of such diverse areas as supernova Type Ia progenitors and binary evolution, and they are predicted to be the strongest gravitational wave sources in the LISA band. Because (ultra)compact binaries show up in light curves with variations on timescales of the orbital period (e.g., due to eclipses or tidal deformation of the components), ZTF is well suited to identify (ultra)compact binaries in a homogeneous way. We expect that the majority of the periodic objects will be typical pulsating stars like Delta Scuti pulsators. The key will be to find the needle in the haystack and select the (ultra)compact systems from the bulk of pulsating stars. A combination of color-selection, proper motions and distances will allow us to distinguish between the bulk of pulsators with potential (ultra)compact binaries, like double white dwarfs, cataclysmic variables (CVs) or hot subdwarf binaries. Among the most numerous CVs will be the large amplitude eruptions of the oldest, lowest mass transfer dwarf novae, allowing a study of the CV graveyard. The large number ($\approx$1000s) of expected (ultra)compact binaries discovered by ZTF will allow us to provide an empirical space density for different types of compact post-common envelope binaries in the Galaxy. The expected large sample will challenge common envelope and binary evolution theories (e.g., predicted vs. observed orbital period and component mass distributions).

Multicolor light curves of eclipsing binary stars allow us to study their stellar parameters in great detail. The duration, depth and shape of the eclipses allows us to determine the relative stellar radii and temperature ratio. We will systematically search the ZTF light curves for eclipsing systems, and use the ZTF $g$, $r$, and $i$ light curves, combined with colors and distances, to determine the system parameters of all eclipsing binaries observed by ZTF. The size of the sample allows us to systematically study the populations of different binary stars. Specifically, we can measure the space density and properties of binary systems that experienced stable or unstable mass transfer. The eclipsing binary sample should also contain rare eclipsing systems. Examples are EL CVn binary stars, eclipsing brown dwarfs, and eclipsing WD systems. In addition, ZTF data also allows us to detect or set an upper limit to the rate of planets around white dwarfs \citep{Agol2011} and large planets around M-dwarf stars \citep[e.g.][]{Bayliss2018}.

Be stars are extreme rotating main-sequence objects that at least once show $H\alpha$ emission line in the spectrum.  Moreover, they are also known as photometric variables.  In a sample of 289 Be stars, \citet{hubert1998} reported that nearly half of them show some photometric variability and, based on the Hipparcos catalog \citep{1997ESASP1200.....E}, an almost entire sample of early Be stars are variables.  A diversity of variability can be found among the Be stars, including non-radial pulsation, intermediate periodicity, long-term variation, semi-regular outburst and outburst variation \citep{2017AJ....153..252L}.  The origin of these variabilities, or so-called the Be phenomenon, remains elusive.  One possible mechanism is the instability of the accretion disk \citep{rivinius2013}.  However, the main challenge in the Be phenomenon is their wide variety of variabilities which required high cadence sampling rate with monitoring over years.  Therefore, most studies of Be phenomenon are on the bright end ($r < 13.5$~mag). Given its combination of ultra wide-field, high cadence and multi-year observations, ZTF provides an opportunity to investigate the Be phenomenon for faint Be stars ($r > 13.5$~mag), especially those found by IPHAS \citep{2015MNRAS.446..274R}, as well as the recent Be candidates selected by PTF \citep{Yu2018} and LAMOST \citep{2015RAA....15.1325L}.  


\section{Small Solar System bodies}
Small Solar System bodies are remnants of the formation stage of the Solar System. They encompass all comets and asteroids, Trojans, Centaurs, near-Earth objects (NEOs) and trans-Neptunian objects. Studies of small bodies contribute to the understanding of several fundamental questions in planetary science, such as the composition of the proto-planetary disk, the evolutionary history of the  Solar System, as well as the transportation and distribution of water and organic materials in the Solar System. Time domain studies of small bodies include their discovery, behavior monitoring, and the detection and rapid follow-up of transient events. The advent of all-sky surveys such as Pan-STARRS, the Catalina Sky Survey, LINEAR, and NEAT have resulted in more small body discoveries at increasingly larger distances from the Sun \citep{Galache2015, Meech2017}. Among current surveys (c.f. \citep{Jedicke2015}), the Zwicky Transient Facility will provide a combination of broad and fast coverage. It will also serve as a precursor to small body observation with LSST \citep{Schwamb2018}, testing the piggyback mode of NEO discovery and operations of mini-surveys.

\subsection{Discovery}
Survey and discovery of NEOs is the critical first step for hazard assessment as well as scientific research. The cumulative efforts of the past few decades have discovered $>95\%$ of kilometer-sized NEOs; however, it is estimated that the coverage of smaller NEOs is less than $10\%$ complete below 200 meters and less than $2\%$ complete below 100 meters \citep[e.g.][]{2015aste.book..795J}. Events like the 2013 Chelyabinsk impact \citep{2013Natur.503..238B} clearly demonstrated the hazard posed by small asteroids. Such asteroids are typically faint and only become visible when they approach the Earth, at which time the ``trailing loss'' effect begins to occur \citep[e.g.][]{2015Icar..257..302H}. The trailing loss presents a challenge for conventional moving object detection algorithms which are tuned to detect point-like sources. For a typical NEO geocentric velocity of 20~km s$^{-1}$, a survey resolution of 1~arcsec, and an exposure time of 30~s, the geocentric distance that trailing loss starts to become significant is about 0.1~AU.

Part of the ZTF NEO discovery effort will be built on the exploratory research done by \citet{2017PASP..129c4402W}, who developed and optimized a pipeline for the detection of trailed NEOs in real-time PTF images. This pipeline is being optimized to enable effective real-time detection of trailed NEOs in ZTF images. ZTF has also deployed a pipeline dedicated to the detection of point-like moving objects \citep{2017PASP..129a4002M} in order to cover the non-trailed moving objects.

The ZTF NEO discovery effort is mainly piggybacked to other surveys, with the exception of the Twilight Survey, a mini-survey that is designed to repeatedly survey the regions with small solar elongation. The goal is to explore interesting phenomena that happen inside the Earth's orbit, including various Sun-approaching comets \citep[e.g.][]{2010AJ....139..926K, 2014ApJ...796...83Y, 2015ApJ...813...73H}; the asteroid population that is predicted to be thermally disrupted \citep{2016Natur.530..303G}; the poorly understood population of Earth Trojans and temporarily-captured natural satellites \citep{Connors2011,Bolin2014}, and NEOs approaching the Earth from the direction of the Sun like the Chelyabinsk event .

All astrometric measurements of trailed and non-trailed moving objects will be submitted to the Minor Planet Center (MPC) in the new Astrometry Data Exchange Standard (ADES) format\footnote{\url{https://minorplanetcenter.net/iau/info/ADES.html}}. The MPC will serve as the liaison for  international follow-up observers. ZTF will use its self-follow-up mode on the P48 system to confirm both ZTF-discovered NEOs and to participate in clearing the MPC's NEO Confirmation Page.

\subsection{Characterization}
\begin{figure}
\includegraphics[width=\columnwidth]{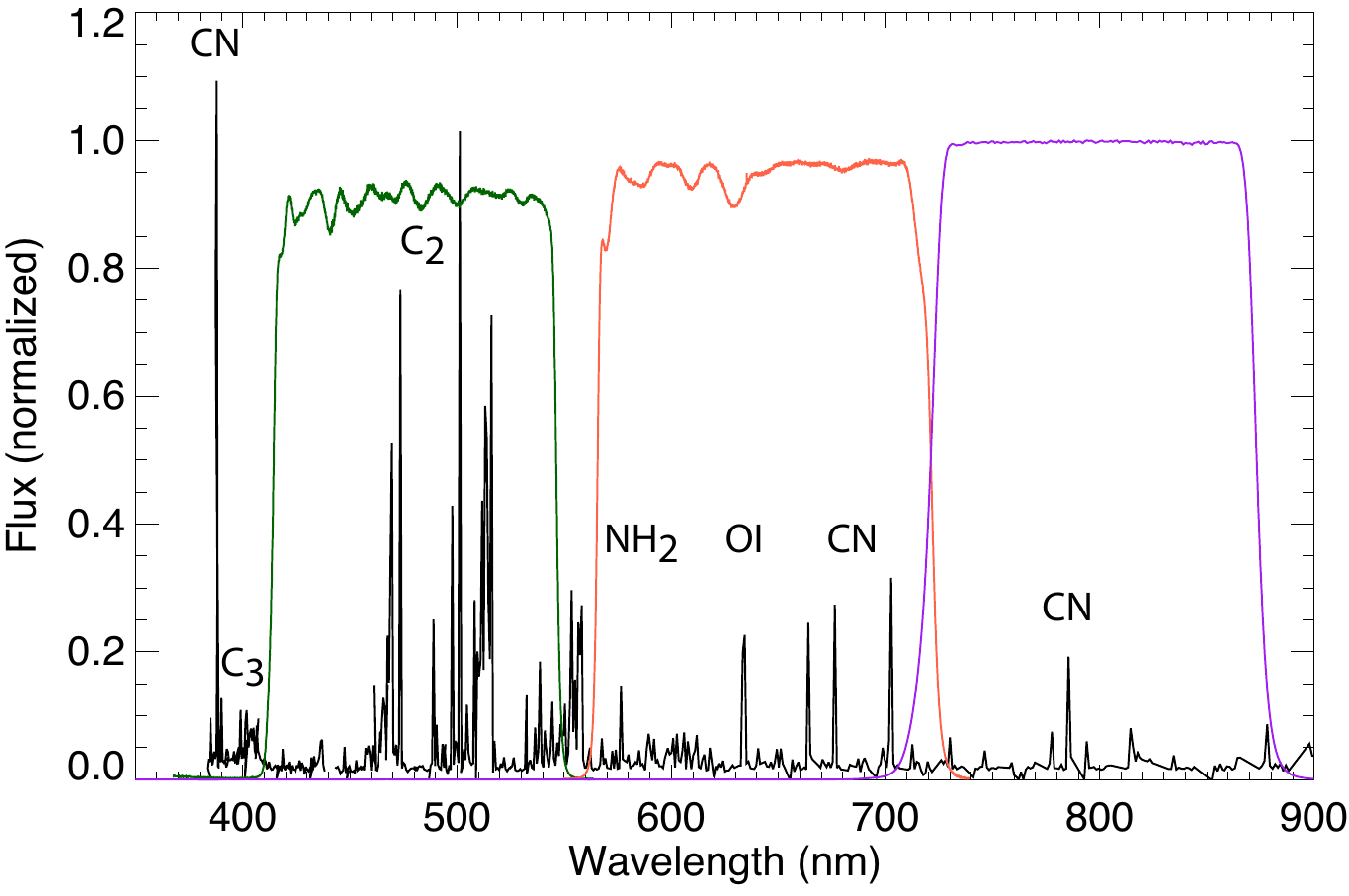}
\caption{Using filter imaging ZTF will be able to monitor the gas and dust content of cometary comae. Here we show the transmission of the ZTF $g$ (green, left), $r$ (red, center), and $i$ (purple, right) filters, superimposed on the spectrum of gas-rich comet 122P/De Vico \citep{Cochran2002}. The $g$ filter contains the emission of C$_2$ molecules, whereas the other filters are mostly free of cometary emission lines. The signal in the $r$ and $i$ filters will come mostly from sunlight reflected by dust surrounding the nucleus.}
\label{CometSpectrum}
\end{figure}

Cometary activity varies with heliocentric distance.  Upon approach to the sun, the temperature of the surface and immediate subsurface is raised, causing ices to sublimate, depending on their proximity to the surface and composition \citep{Meech2004, Prialnik2004}.  Therefore, a comet's coma may change composition as it orbits the sun.  In addition, the seasonal context of the nucleus affects which surface areas receive sunlight, and provides another means for coma variability.  Variations of coma brightness (i.e., activity) and color (i.e., composition) thus provides a means of exploring heterogeneities of comet nuclei.  We predict that ZTF can detect at least 30 comets per night (V < 21), and many of those comets will be observable for periods longer than a year. Depending on sky conditions, comet brightness, and background, we anticipate a photometric accuracy of 0.1 mag. 

Our photometry methods allow us to use ZTF to systematically produce accurate comet light curves. As is shown in Fig.~\ref{CometSpectrum}, images acquired in different filters can be used as a diagnostic of the comet's dust-to-gas ratio over time; the $g$ filter contains bright emission lines of C$_2$ molecules, whereas the $r$ and $i$ bandpasses are mostly free of cometary emission features and thus sample sunlight reflected by dust in the coma.

Asteroid light curves can be used to measure several fundamental properties for asteroids, such as shape, spin status and taxonomy. The statistics of asteroid rotations can help to understand how their evolution is affected by mutual collisions, gravitational perturbations of planets, and the YORP effect. Phase functions can be used to determine asteroid taxonomy. Combining rotation period and taxonomy, we are also able to study the spin-rate limits for different type asteroids, which is a proxy of asteroid bulk density \citep{Chang2015}. We expect to collect $\sim$100,000 asteroid light curves per year from which we will be able to derive rotation periods and phase functions.  

Most asteroids are gravitationally bounded aggregations (`rubble-piles'). It is thought that rubble-pile asteroids cannot have rotation periods less than a critical limit, (i.e., the 2.2-hr spin-rate limit \citep{Harris1996}) or they will disintegrate. However, it has been found that a small number of asteroids have rotation periods shorter than this limit, implying that they may have different structure from the average asteroid. PTF has discovered 3 of the 6 super-fast rotators (SFRs) known to date \citep[][and the references therein]{Chang2017,Waszczak2015}. However, the detection rate is still too low to place a meaningful constraint to the SFR population. With its large sky coverage, ZTF can improve our knowledge to the SFR population.

The binarity of asteroids probes the Solar System collisional evolution
\citep[e.g.][]{Pravec2010} and measures the dynamical mass of asteroids.
A possible method for searching for binary asteroids and measuring their mass
is via looking for astrometric variations from a pure Keplerian orbit.
This method is currently being tested on PTF data (Polishook \& Ofek, in prep.).
Given the large number of asteroid images we expect to acquire with ZTF,
this simple method may enable us to find most binary asteroids (and Kuiper Belt Objects) in
the Solar System.

\subsection{Transient Events}
The systematic, high-cadence monitoring of small Solar System bodies will provide a baseline that allows ZTF to find transient events such as cometary outbursts and fragmentation events \citep{Ye2015,Ishiguro2016}, collisions between asteroids \citep{Snodgrass2010, Bodewits2011}, unexpected or irregular activity in asteroids \citep{Jewitt2012, Waszczak2013, 2017AJ....153..207Y} and Centaurs \citep{Jewitt2009}. The cadence of the ZTF observations will allow us to evaluate the frequency of these events. The early discovery of such transient events enables rapid follow-up observations which are critical for the characterization of these events, because the ejected material quickly sublimates or dissipates away. 

\section{Astroinformatics \& Astrostatistics}



%
ZTF is well positioned to enable the fields of astroinformatics and astrostatistics make significant strides through the development and testing of novel computational and statistical methodologies related to large data sets.
These new algorithms and techniques will not only be useful for analyzing ZTF data, but will also provide a ready-to-use analysis toolkit for data from future surveys like LSST. Indeed, the anticipation of ZTF data has already led to the development of new data-processing pipelines by {\it IPAC} \citep{tmp_Masci:18:ZTFDataSystem} and the implementation of the Kafka system \citep{tmp_Patterson:18:ZTFAlertDistribution}.

ZTF data will necessitate new statistical methodologies and machine learning algorithms specifically designed for astronomy and astrophysics. In particular, methods for time series analysis and populations studies that allow testing and comparison of physical models are needed. Additionally, reliable classification algorithms will be needed to properly perform scientific inference. Finally, model comparisons using modern statistical methods will be fundamental in ruling out physical models in light of complex ZTF data.

As large data sets from projects such as ATLAS, ASAS-SN, CRTS, Gaia, JWST, LSST, Pan-STARSS, etc. continue to become available into the 2020s and 2030s, there will be a demand for statistical and computational methodologies which not only handle large amounts of data but also extract the most information possible. ZTF is unique in that, combined with its follow-up network, it will provide one of the first opportunities to develop and test methods for large datasets that have significant temporal information.
\subsection{Population Studies}
ZTF will discover many faint and fast-evolving transients, allowing a search for subpopulations and tests of proposed physical mechanisms behind these phenomena  (Section~\ref{sec:SESN}). ZTF will also provide us with an unbiased and complete sample of SNe Ia within $z<0.1$, which in turn will allow us to investigate the effects of host environments (Section~\ref{sec:SNIa}). In both of these cases, studying the population of light curve data is key to identifying patterns and subpopulations. Thus, modern cluster-finding algorithms for time series data are needed to analyze the light curves of these transients.

Methods for time series clustering exist in other research areas such as statistics, finance, medicine, and economics, but not all of these new methods have made it to astronomy \citep[for a review of time series clustering methods, see][]{TSclustering2015}. Recurring challenges for time series clustering methods include how to deal with missing data and the reliance on mathematical distance measures. Recently, new methods have been proposed that overcome these challenges  \citep[e.g.][]{Wang2006} and these could be useful for the analysis of ZTF light curve population studies. At the same time, astronomical time series of transient sources bring new challenges to the table--- for example, uncertainties in distance to the source and how to account for reddening due to dust.

The curve data sets of specific populations (such as SE~SNe and SNe~Ia) will present an opportunity for astroinformatics and astrostatistics to bring in methods from other research areas to make discoveries, and to build upon and tailor these techniques for astronomy. 

\subsection{Classification}

Classification of one form or another will lie at the heart of solving many of the ZTF science cases. The sheer number of sources and alerts from ZTF will necessitate  reliable classification tasks with little human intervention. Over the past twenty years or so, the proliferation of algorithms in the field now called machine learning has led to the creation of a powerful toolbox of methods that can perform classification in a large variety of different contexts. The challenge here lies in the structure of both the problems to be addressed and the data itself. 

In some instances, correctly predicting the  source or alert type may be all that is required, but the ultimate goal of most classification tasks in the context of ZTF is scientific inference. For example, when classes of sources are identified with the goal of performing the population studies mentioned above, biases within the classified data set must be carefully assessed and propagated through to the inference stage. Achieving the latter is not always straightforward with many of the newer deep learning methods, although recent developments related to (local) interpretability \citep[e.g.][]{ribeiro2016should,2016arXiv160605685K} and probabilistic machine learning \citep[e.g.][]{tran2016edward,ghahramani2015probabilistic} may be able to either shed light on these biases or incorporate them directly into the subsequent modelling tasks. 

In contrast to data sources often considered in machine learning contexts,  the data derived from the ZTF survey will be very heterogeneous, unevenly sampled, and subject to occasionally catastrophic outliers. Moreover, the data will include variable uncertainties. Recent work shows a range of different approaches for dealing with such issues. Promising results are found through methods such as recurrent neural networks \citep{naul2018recurrent}, convolutional neural networks trained on two-dimensional representations of light curves \citep{2017arXiv170906257M}, and the use of deep neural networks for phenomenological discovery of variable star classes \citep{2018arXiv180402132H}.

The peculiarities of the ZTF data ---  combined with the requirement to classify sources and subsequently perform inference --- implies that standard techniques may not deliver the performance necessary to answer the scientific questions in this paper. 
However, these constraints also present an opportunity to develop new classification methods that can be carried forward to future surveys that share the same challenges. 

In addition to the real-bogus separation covered in \citet{tmp_Mahabal:18:ZTFMachineLearning},
we identify three objectives for classification with ZTF data: classification for follow-up, classification for scientific inference, and finding the unexpected. Each objective presents its own challenges. Below, we show where recent research from other domains could be usefully applied to ZTF data or where new methods must be developed to overcome these challenges.

With a projected one million alerts per night, ZTF will produce a large number of transients. Because observing time is scarce and expensive, there is significant impetus to optimize which sources should be followed up with other facilities, and how soon  they should be observed in order to maximize scientific output. Any algorithm must be capable of dealing with a continuously changing data set on top of the heterogeneity mentioned above, and be able to update predictions almost on-the-fly, for example as part of an alert brokering system \citep[e.g. ANTARES:][]{narayan2018machine}. There are a wealth of methods related to Active Learning (i.e. learning with feedback; \citealt{2018arXiv180403765I,8285192,7929964,7372134}) and Online Learning (i.e. learning from evolving data streams; \citealt{aggarwal2007data,Nguyen2015}) that can be tested with ZTF in order to both enable follow-up studies as well as prepare for future surveys.
These approaches may be combined with other techniques \citep[e.g. probabilistic forecasting techniques, see][for an example]{kuznetsov2015learning} in order to estimate when a future follow-up observation should be taken. Only by optimizing  the information gathered about the source can competing models be rigorously tested. Owing to the data volume, methods deployed on clusters, GPUs, and other multi-processing hardware will find increasing use (e.g. \cite{2015arXiv151202831G,8285225}). 

As mentioned above, many science cases rely heavily on identifying a complete subset of the relevant source type (e.g. for SN Ia cosmology, see also Section \ref{sec:SNIa}). A key challenge in identifying a complete subset with machine learning is the dearth of complete, applicable training data sets for which the truth is known. One option is to simulate data sets for a given instrument, but the correctness of the derived prediction depends crucially on the assumption that the simulated training data matches the real test data exactly. Many algorithms, especially more modern deep learning frameworks, tend to be extremely vulnerable to mismatches between training and test data \citep[e.g.][]{evtimov2017robust}. Recent advances in the field of transfer learning may make it possible to use training data sets generated with other surveys and use the information in them to generate accurate classifications for ZTF sources \citep{aswolinskiy2017unsupervised,2017arXiv170607446G}.

In addition, the field of probabilistic machine learning has recently received much attention from within the computer science community. Different domains have seen an emergence of time series methods within a probabilistic framework, within machine learning (e.g. to model motion capture data \citep{2018arXiv180206765A}, housing prices \citep{2017arXiv170709380G} or homelessness \citep{ren2017}). These methods align well with the science goals of ZTF, and will allow for direct propagation of uncertainties and biases into the resulting astronomical inferences.
However, even with transfer learning, there might not exist a complete, unbiased training data set for ZTF, in which case unsupervised methods (see below) may present a better solution.

Serendipity has traditionally been a strong component of astronomical discovery, especially when the instrument in question opens up a new part of parameter space. In addition, some science cases have no strong prior on the number or types of classes. Much of the work within the machine learning community has focused on supervised machine learning. However, finding the unknown with traditional supervised methods is exceedingly difficult. Here, new approaches to unsupervised machine learning may help us discover unknown transients and new source classes. In particular, the recent development of methods for the classification of sparse, irregularly sampled time series based on Gaussian Processes \citep[e.g.][]{li2016scalable,ghassemi2015multivariate}, deep learning \citep[e.g.][]{lipton2016directly,che2018recurrent} and time series clustering have shown promise across multiple domains.



\section{Summary}

In this paper we have summarized the main science drivers that led to the
Zwicky Transient Facility (ZTF) consisting of a 47\,deg$^2$ imager on the Palomar 1.2-m Oschin (Schmidt) telescope and a low resolution spectrometer (the Spectral Energy Distribution Machine or SEDM) on the Palomar 60-inch telescope. From cometary outbursts and asteroid collisions to infant SNe and failed GRB jets, from Be stars and ultracompact binaries to interactions with (supermassive) black holes, populations of transient and variable astrophysical sources can now be studied in great detail. Although comprehensive, this list is not exhaustive and the ZTF public alert stream offers the community ample opportunity to make their own discoveries.

The rates of potential discoveries -- supernovae and asteroids every night, a changing-look AGN every week, a TDE every month, a gravitationally-lensed SN every quarter, and maybe a few unexpected transients a year -- with each easily followed-up spectroscopically by a 5--10 m class telescope (and smaller for photometry) take us into the new territory of deciding each night what the most interesting sources currently are and whether the previous night's sources still merit continued attention. This problem will become acute once even more powerful facilities come online, in particular, LSST.  Even with our considerable follow up resources we are not in a position to follow {\it all} transients identified with ZTF. We call this problem the conundrum of abundance. 

To start with abundance is good, particularly for astronomy that depends primarily on
photometric data. For instance, consider the search for rare types of variable stars (e.g., 
double degenerates with very short periods). The larger the data set, the higher the chance
of discovery. Large data sets also allow the extraction of huge samples of ordinary phenomena
(e.g., RR Lyrae stars) and large samples could lead to identification of finer sub-classes. 

The conundrum of abundance is a problem for transient object science in which follow-up
is needed.\footnote{For well behaved transients such as Type Ia, a purely photometric
approach using photometric redshifts for host galaxies is certainly feasible.} A million alerts per night ensures that the sky is always saturated 
with follow-up targets (from NEOs to TDEs), enabling us to select and focus on the ones most likely to yield the highest value results (depending on a group's area of interest).
The solution is to sharply define science
programs that can be undertaken with existing facilities. This would mean
designing filters that efficiently find desired transients whilst suppressing
false positives. In fact, in this respect, we have already achieved good performance in the area of SLSNe, TDEs and relativistic transients. Bright transients will
always remain interesting. Given the success of the SEDM we advocate similar
low resolution spectrographs for 2-m class telescopes. 


Finally, the night sky is finite in extent and with multiple facilities scanning the same regions every night, there is clearly scope for synergies to optimize scientific discovery. A key component of this has to be adequate community infrastructure and coordination to support the real-time distribution, characterization, and classification of million of alerts per night from surveys (let alone followup observations of interesting sources), as well as systematic searches of archives of billions of time series. ZTF is clearly a pathfinder for some of this and coordinated efforts across surveys drawing on it will create the basis for LSST and other time domain surveys of the next decade and beyond.

\acknowledgments
Based on observations obtained with the Samuel Oschin 48-inch Telescope and the 60-inch Telescope at the Palomar Observatory as part of the Zwicky Transient Facility project, a scientific collaboration among the California Institute of Technology, the Oskar Klein Centre, the Weizmann Institute of Science, the University of Maryland, the University of Washington, Deutsches Elektronen-Synchrotron, the University of Wisconsin-Milwaukee, and the TANGO Program of the University System of Taiwan. Further support is provided by the U.S.\ National Science Foundation under Grant No.\ AST-1440341.

J.~Sollerman acknowledges support from the Knut and Alice Wallenberg Foundation.

E.~Ofek is grateful for support by  a grant from the Israeli Ministry of Science,  ISF, Minerva, BSF, BSF transformative program, and  the I-CORE Program of the Planning  and Budgeting Committee and The Israel Science Foundation (grant No 1829/12).

A.~Gal-Yam is supported by the EU via ERC grant No. 725161, the Quantum Universe I-Core program, the ISF, the BSF Transformative program and by a Kimmel award. 

S.~Gezari is supported in part by NSF CAREER grant 1454816 and NSF AAG grant 1616566.

C.-K.~Chang, W.-H.~Ip, C.-D.~Lee, Z.-Y.~Lin, C.-C.~Ngeow and P.-C.~Yu thank the funding from Ministry of Science and Technology (Taiwan) under grant 104-2923-M-008-004-MY5, 104-2112-M-008-014-MY3, 105-2112-M-008-002-MY3, 106-2811-M-008-081 and 106-2112-M-008-007.

M. Bulla and A. Goobar acknowledge support from the Swedish Research Council (Vetenskapsr\aa det) and the Swedish National Space Board.

E. Bellm, B. Bolin, A. Connolly, V. Z. Golkhou, D. Huppenkothen, Z. Ivezi\'{c}, L. Jones, M. Juric, and M. Patterson 
acknowledge support from the University of Washington College of Arts and Sciences, Department of Astronomy, and the DIRAC Institute. University of Washington's DIRAC Institute is supported through generous gifts from the Charles and Lisa Simonyi Fund for Arts and Sciences, and the Washington Research Foundation. M.~Juric and A.~Connolly acknowledge the support of the Washington Research Foundation Data Science Term Chair fund, and the UW Provost's Initiative in Data-Intensive Discovery.

E. Bellm, A. Connolly, Z. Ivezi\'{c}, L. Jones, M. Juric, and M. Patterson 
acknowledge support from the Large Synoptic Survey Telescope, which is supported in part by the National Science Foundation through
Cooperative Agreement 1258333 managed by the Association of Universities for Research in Astronomy
(AURA), and the Department of Energy under Contract No. DE-AC02-76SF00515 with the SLAC National
Accelerator Laboratory. Additional LSST funding comes from private donations, grants to universities,
and in-kind support from LSSTC Institutional Members.

B.T. Bolin acknowledges funding for the Asteroid Institute program provided by B612 Foundation, W.K. Bowes Jr. Foundation, P. Rawls Family Fund and two anonymous donors in addition to general support from the B612 Founding Circle.

M.T. Soumagnac acknowledges support by a grant from IMOS/ISA, the Ilan Ramon fellowship from the Israel Ministry of Science and Technology and the Benoziyo center for Astrophysics at the Weizmann Institute of Science.

A.A.\ Miller is funded by the Large Synoptic Survey Telescope Corporation in support of
the Data Science Fellowship Program.

J.~Bauer, T.~Farnham, and M.~Kelley gratefully acknowledge the NASA/University of Maryland/MPC Augmentation through the NASA Planetary Data System Cooperative Agreement NNX16AB16A.

M. M. Kasliwal and Q.-Z. Ye acknowledge support by the GROWTH (Global Relay of Observatories Watching Transients Happen) project funded by the National Science Foundation PIRE (Partnership in International Research and Education) program under Grant No 1545949.

A.~A.\ Mahabal acknowledges support from the following grants: NSF AST-1749235, NSF-1640818 and NASA 16-ADAP16-0232.

M. W. Coughlin is supported by the David and Ellen Lee Postdoctoral Fellowship at the California Institute of Technology.

S. Ghosh acknowledges the NSF Award PHY-1607585.

M. Rigault acknowledges funding from the European Research Council (ERC) under the European Union's Horizon 2020 research and innovation programme (grant agreement no. 759194 - USNAC).

\facilities{PO:1.2m, PO:1.5m}

\bibliographystyle{yahapj}
\bibliography{main}





\end{CJK*}
\end{document}